\documentclass[10pt,preprint]{aastex}

\usepackage{epsfig,graphicx,amssymb}

\newcommand{\Swift}{{\em Swift}}
\newcommand{\swift}{\textit{Swift}} 

\newcommand{\chandra}{\textit{Chandra}}

\newcommand{\xmm}{\textit{XMM-Newton}}

\newcommand{\pnoid}{\mbox{$P_{\rm no-id}$}}

\def\nhtt{\mbox{$N_{\rm H,22}$}}
\def\nh{\mbox{$N_{\rm H}$}}
\def\deg{\hbox{$^\circ$}}
\def\arcdeg{\hbox{$^\circ$}}
\def\arcmin{\hbox{$^\prime$}}
\def\arcsec{\hbox{$^{\prime\prime}$}}
\def\h{\hbox{$^{\rm h}$}}
\def\m{\hbox{$^{\rm m}$}}
\def\s{\hbox{$^{\rm s}$}}

\def\simlt{\mathrel{\hbox{\rlap{\hbox{\lower4pt\hbox{$\sim$}}}\hbox{$<$}}}}
  \def\simgt{\mathrel{\hbox{\rlap{\hbox{\lower4pt\hbox{$\sim$}}}\hbox{$>$}}}}

\def\unde{Undetermined}
\def\pchance{\mbox{$P_{\rm chance}$}}
\def\done{}

\newcommand{\xray}{\mbox{X-ray}}

\newcommand{\rosat}{{\em ROSAT}}
\newcommand{\erosita}{{\em eROSITA}}

\newcommand{\perval}[2]{{#1\mbox{$^{#2}$}}}
\newcommand{\msun}{\mbox{$M_\odot$}}

\newcommand{\persec}{\perval{\rm s}{-1}\/}
\newcommand{\perksec}{\perval{\rm ks}{-1}\/}
\newcommand{\cpks}{\mbox{counts}\,\perksec}
\newcommand{\percm}{\mbox{$\cm^{-2}$}}

\newcommand{\ppm}{\mbox{$\pm$}}
\newcommand{\ee}[1]{\mbox{$10^{#1}$}}
\newcommand{\tee}[1]{\mbox{$\times 10^{#1}$}}

\newcommand{\erg}{\mbox{$\rm\,erg$}\/}
\newcommand{\cm}{\mbox{$\rm\,cm$}}
\newcommand{\ksec}{\mbox{$\rm\,ks$}}

\newcommand{\fxfopt}{\mbox{$F_{\rm X}/F_{\rm opt}$}}

\newcommand{\kteff}{\mbox{$kT_{\rm eff}$}}

\newcommand{\ud}[2]{\mbox{$^{+ #1}_{- #2}$}}
\newcommand{\cgsflux}{\erg\,\percm\,\persec}
\newcommand{\cgslum}{\erg\,\persec}

\newcommand{\tnm}[1]{\tablenotemark{#1}}
\newcommand{\pobs}[1]{\textit{#1}}

\newcommand{\notanins}{Not an INS}

\newcommand{\obsid}{observation ID}

\begin{document}


\title{A Limit on the Number of Isolated Neutron Stars
  Detected in the ROSAT All-Sky Survey Bright Source Catalog}

\author{Monica L. Turner\altaffilmark{1}, 
Robert E. Rutledge\altaffilmark{1}, 
\altaffiltext{1}{Department of Physics, McGill University,
  3600 rue University, Montreal, QC, H3A 2T8, Canada;
  turnerm@physics.mcgill.ca, rutledge@physics.mcgill.ca} 
Ryan Letcavage\altaffilmark{2}, 
Andrew  S. H. Shevchuk\altaffilmark{2}, 
Derek B. Fox\altaffilmark{2}
\altaffiltext{2}{Department of Astronomy \& Astrophysics,
  525 Davey Laboratory, Pennsylvania State University, 
  University Park, PA 16802, USA; 
  rjl5011@psu.edu,  ahs148@psu.edu, dfox@astro.psu.edu}
}

\slugcomment{}
\shorttitle{Limit}
\shortauthors{}


\begin{abstract}

Using new and archival observations made with the \swift\ satellite 
and other facilities, we examine 147 \xray\ sources selected from
the \rosat\ All-Sky-Survey Bright Source Catalog (RASS/BSC)
to produce a new limit on the number of isolated neutron stars 
(INS) in the RASS/BSC, the most constraining such limit to-date.
Independent of \xray\ spectrum and variability, the number of INSs is
$\leq$48 (90\% confidence).  Restricting attention to soft ($kT_{\rm
  eff}<200$\,eV), non-variable \xray\ sources -- as in a previous
study -- yields an all-sky limit of $\leq$31 INSs.  In the course of
our analysis, we identify five new high-quality INS candidates for
targeted follow-up observations.
A future all-sky \xray\ survey with \erosita, or another mission with
similar capabilities, can be expected to increase the detected
population of \xray-discovered INSs from the 8 to 50 in the BSC, to
(for a disk population) 240 to 1500, which will enable a more detailed
study of neutron star population models. 

\end{abstract}

.

\maketitle

\section{Introduction}
\label{sec:intro}

The \xray\ source class of isolated neutron stars (INSs) are
observationally defined by their high \xray\ to optical flux ratio,
$\fxfopt\simgt 10^4$ \citep{treves00}. At discovery, these objects 
do not exhibit radio pulsations and are not associated with supernova 
remnants.

The study of this phenomenological class ultimately promises a more
complete picture of neutron star properties and evolution.  Following
formation in a supernova explosion (SN), the ``standard'' neutron
star (that is, one with a core composed of beta-equilibrium nuclear
matter) evolves thermally, cooling to a temperature $T\sim\ee{6}$\,
after $t\sim\ee{6}$\,yr \citep{page04}. The cooling is modestly
affected by the presence of strong ($B\sim\ee{15}$\,G) magnetic fields
\citep{arras04}.  During this period, if the neutron star is emitting
isotropically in the \xray\ band, detection and study of the source
will be a straightforward issue of improving the flux sensitivity of
\xray\ instrumentation.  This simple picture could be altered if
neutron stars, at birth, have strong toroidal magnetic fields, which
may significantly modify the isotropy of their early emission
\citep{page07}.  

Over 1500 radio pulsars have been discovered \citep{manchester05}; and
a few hundred low-mass-X-ray binaries (LMXBs) and high mass \xray
binaries (HMXBs) are catalogued \citep{liu06,liu07}.  However, star
formation models predict $N_{\rm MW}\sim\ee{9}$ neutron stars have
formed within the Milky Way \citep{timmes96}, implying that the vast
majority of stellar evolution remnants have not yet been observed.

While one can attempt to infer the natal properties of neutron stars
observed as radio pulsars and \xray\ binaries, selection effects are
present in both populations.  Radio pulsars exhibit significant
surface dipolar magnetic fields ($B\sim\ee{8}-\ee{13}$\,G; 
\citealt{camilo94,camilo00}); the distribution of magnetic field
strength and geometry at birth, and their subsequent time evolution,
are not strongly observationally constrained; it may be that only a
fraction of neutron stars are born with such magnetic fields, implying
that relatively few NSs are ever observable as radio pulsars.

The uncertain emission mechanism(s) of radio pulsars have made it
challenging to quantify the observational selection effects which
impact their detection in radio surveys.  Nonetheless, recent analyses
modelling this class concludes that the properties of known pulsars and
pulsar surveys, including best estimates of the above selection
effects, can account for all neutron stars observed \citep{faucher06}.

Observation and characterization of the INS population (which, in the
present work, will include all neutron stars which are first
identified via their high \fxfopt) provides a means to more fully
characterize the neutron star population of the Galaxy, without
selections related to magnetic field strength, geometry, evolution,
and the various phases (and consequences) of binary stellar evolution.

Modelling of the INS population is, however, susceptible to theoretical
uncertainties in neutron star cooling mechanisms, including the
possibility of enhanced neutrino emissivity due to hyperon condensates
\citep{page04}.  Determining the ages (to $\simlt$50\%) of individual
isolated thermally-emitting neutron stars, independent of the observed
surface temperature, is currently only possible for objects with
precision parallax and proper motion measurements (e.g.,
\citealt{kaplan02b,kaplan07}), which allow their space velocity to be
back-extrapolated to a likely birthplace in an OB-association.

In the absence of accurate ages for individual INSs, comparing
individual INS properties with cooling models is effectively an
interpretive act rather than a confrontation of theory with data.  To
test these models, therefore, will ultimately require a sufficient
number of observed INSs such that models of their birth properties,
evolution, and dynamics can be challenged to reproduce the observed
population, as is now usefully done for the radio pulsar population
\citep{cordes98,arzoumanian02,gonthier04,faucher06}.

INSs evolve in the absence of influence of a companion (such as
accretion of $\sim$0.1-0.3 \msun over the lifetime of a low-mass X-ray
binary), and so offer an opportunity to study the physics of compact
objects directly -- their formation, dynamics, thermal evolution and
magnetic field evolution.  Initial estimates \citep{blaes93} for the
number of INSs which would be detected in the \rosat\ All-Sky-Survey
Bright Source Catalog (RASS/BSC \citep{voges99}) overestimated the
number which were eventually observed, by a factor of $\sim$100 (as
discussed previously; \cite[R03 hereafter]{rutledge03}).
This overestimation was due to the fact that the observable
population's luminosity was thought to be dominated by simple
Bondi-Hoyle accretion, which scales approximately with the neutron
star velocity $\propto v^{-3}$.  However, the velocity distribution for
radio pulsars was later found to underestimate the typical
velocity by a significant factor, which subsequently altered the
conclusions by decreasing both the luminosity and the number of
detectable INSs.  Later estimates, which employ results of MHD
simulations of Bondi-Hoyle accretion, found that the modified Bondi
formula dramatically suppresses the observable population even
further, to the extent that none are expected detected above RASS/BSC
sensitivity \citep{perna03}.  Thus, the interpretation of INSs
observed from the RASS/BSC has been as post-natal cooling from neutron
stars
\citep{popov00b,popov01,popov03,popov06,popov06b,posselt07,posselt08}.

The present work continues our search for INSs -- \xray-bright sources
with no observed off-band emission at discovery. This follows our
previous work (R03), using essentially the
same selection methods for INS candidates detected by the RASS/BSC.
We have expanded the selection of INS candidates based on \pnoid\ --
the probability that the RASS/BSC X-ray source is not associated with
any off-band counterparts in the catalogs considered (USNO-A2; NVSS
and IRAS) -- from $\pnoid\geq0.90$ to $\pnoid\geq 0.8$.  Doing so
increases the fraction of all INSs detected within the analyzed region
of the RASS/BSC ($\delta\geq -39\deg$, due to the sky-coverage of
NVSS) which pass our selection, from $\sim$20\% (R03) to 30\% (in the
present work); this insures a greater number of INSs detected in the
RASS/BSC end up in our selected INS candidates list.

To find 100\% of the INSs within the RASS/BSC, an
analysis such as the one performed here would be required for all
18806 RASS/BSC X-ray sources.  The present analysis makes use of a
statistical selection which reduces the number of sources to be
analyzed by a factor of x100 (from 18806 RASS/BSC sources, to 147
sources), and will permit identification of almost 1/3 of all INSs
among RASS/BSC sources. To identify 100\% of the INSs in the RASS/BSC
sources, thus will require approximately 100$\times$ the amount of
observations analyzed here, to obtain a factor of 3 more INSs.  Thus,
searches for INSs and classification of their populations using the
techniques described here involve a balance between observational
resources invested (in \swift\ and \chandra\ observations), and the
number of INSs which will be identified.  The size of the sample
analyzed here is a compromise between these two goals, providing a
significant improvement over previous characterization of the INS
population in the RASS/BSC \cite{rutledge03}.

We also report follow-up observations with \swift, as well as archival
observations with \chandra, \xmm, and \rosat/HRI, which permit
localization and association with off-band (optical, IR and UV)
sources. 

As previously, we note that the present analysis produces an
upper-limit not only on objects that follow the strict definition
of an INS - non-radio pulsar, non-magnetar, and non-SNR associated
objects - but of high X-ray/optical flux ratio-selected populations
of {\em all types} (such as anomalous X-ray pulsars (AXPs)).
It should be noted that in our analysis, we may refer to objects as
``INS'' based on their X-ray/optical flux ratio and not on other
properties that may technically exclude it from the phenomenological
class. As an example, one object in our selection sample,
1RXS~J141256.0+792204 (also known as Calvera) was identified as
an isolated compact object of some kind based on its X-ray/optical
flux ratio \citep{rutledge08}. Although more observations are
necessary to determine its classification as either an INS, AXP,
or radio pulsar, we classify it in our analysis as an ``INS''.


\section{Observations \& Analysis}
\label{sec:obs}

The analysis proceeds in three parts: 

\begin{enumerate}

\item Selection of the candidate INS sources from the RASS/BSC and
off-band catalog cross-correlations, to find X-ray sources which are
not likely to be associated with off-band counterparts. 

\item Follow-up observations (with \swift/XRT + UVOT); literature
  search for well-observed and understood counterparts to the RASS/BSC
  sources; and archival analysis of \chandra\ (0.6\arcsec\ systematic
  positional uncertainty\footnote{From the \chandra\ calibration web
  page available at http://cxc.harvard.edu/cal}, 1$\sigma$), \xmm\
  (2\arcsec systematic positional uncertainty\footnote{From \xmm\
  Science Operations Centre, XMM-SOC-CAL-TN-018 (Guainazzi 2008)},
  1$\sigma$), and \rosat/HRI (6.3\arcsec\ systematic positional
  uncertainty\footnote{See
  http://www.mpe.mpg.de/xray/wave/rosat/doc/ruh/node27.php}) which
  permit refined localization of the RASS/BSC X-ray source to search
  for off-band associations. We consider two point sources to be spatially 
  associated if the probability that an as bright or brighter off-band 
  catalogued object should lie as close or closer to the RASS/BSC X-ray source is
  $\leq1\%$;  ; that is, the probability of chance association between
  the RASS/BSC source and the optical source (\pchance) is
  $\leq$\ee{-2}.  For the number of RASS/BSC sources in our sample
  (147), this will produce no more than $\sim$1-2, on average, false
  associations with off-band counterparts.  With regards to literature
  searches: some RASS/BSC sources have multiple X-ray observations in
  which phenomena (such as type-I X-ray bursts) are observed which are
  not observed from INS, while others do not.  We offer references for
  work which determines the object is ``not an INS'', or we otherwise
  conclude that the object is classified as ``undetermined.''

\item Following the same statistical procedure as employed in our
  previous work (R03), we calculate a new upper-limit on the number of
  INSs in the RASS/BSC.

\end{enumerate}

In the following subsections, we provide further detail on each of
these steps.


\subsection{Selecting the RASS/BSC INS Candidate Sample}

We draw candidate sources from the same selection as our previous
analysis (R03), in which a full description of this selection is
found.  In brief, we performed a statistical cross-identification
between the RASS/BSC and the USNO-A2 (optical), IRAS (infrared), and
NVSS (radio) catalogs, quantifying (among other things) the
probability \pnoid\ that each RASS/BSC X-ray source was not associated
with any of the off-band catalog sources. 

To calculate \pnoid\, all sources from USNO-A2 within 75\arcsec\ and 
from IRAS and NVSS within 150\arcsec\ of the RASS/BSC positions were
used to determine a likelihood of association for every X-ray source 
$i$ and off-band source $j$ from catalogue $C$.  This likelihood ratio 
is given by
\begin{equation}
LR_{i,j;C}=\frac{\mbox{exp}\left(-r^2_{i,j}/2\sigma^2_{i,j}\right)}
{\sigma_{i,j}N(>F_j;C)}
\end{equation}
where $r_{i,j}$ is the separation between the X-ray and off-band source;
$\sigma_{i,j}$ is the uncertainty in $r$, and $N(>F_j;C)$ is the number 
of sources in the catalog $C$ with fluxes $F$ greater than off-band source 
$j$.  

Background fields, which are 24 off-source positions set in a 5$\times$5 
grid around each X-ray source position, consist of circles with radii of 
75\arcsec\ separated by 150\arcsec\ for USNO-A2, and circles with radii of 
150\arcsec\ separated by 300\arcsec\ for IRAS and NVSS.  The probability that a
candidate counterpart $j$ is spatially associated with the X-ray source  
and is not a background object is given by 
\begin{equation}
 R_{i,j}(LR_{i,j;C})=\frac{N_{\mbox{src}}(LR_{i,j;C})-N_{\mbox{bkg}}(LR_{i,j;C})}
{N_{\mbox{src}}(LR_{i,j;C})}
\end{equation}
where $N_{\mbox{src}}(LR_{i,j;C})$ is the number of objects in the source 
fields that have a value of $LR$ within some $\delta LR$ of $LR_{i,j;C}$
and $N_{\mbox{bkg}}(LR_{i,j;C})$ is the same but for the background fields.

Thus we can obtain the probability that none of the objects in the field are 
associated with the X-ray source, given by
\begin{equation}
 P_{i,{\rm no-id}}=\frac{\Pi_j(1-R_{i,j})}{K}
\end{equation}
 where $K$ is a normalization factor.  A high value of \pnoid\ for
a RASS/BSC X-ray source means it is unlikely that the RASS/BSC source
is associated with any of the off-band catalog sources.

For the purpose of the analysis, 150 "control" X-ray sources were inserted 
among the source fields, distributed randomly in proportion to the local 
X-ray source density.  Since they should have no detectable off-band 
counterparts, they behave like INSs.  Thus, we are able to evaluate the 
statistical efficiency of our selection, based on the number of control 
sources that remain above our $P_{no-id}$ cut-off.  

We expand upon the earlier selection in the following two ways: 

\begin{enumerate}

\item We now include RASS/BSC sources with \pnoid$>$0.8
  (vs.\ \pnoid$>$0.9 in our previous work); in addition to increasing
  the number of candidate sources to be investigated, this increases
  the number of control sources which end up in our selection.
  Previously 29 of 150 (19\%) control sources ended up in our
  selection; here, 46 of 150 (31\%).  Because control sources have the
  same behavior expected from INSs (i.e., have no counterparts in the
  off-band catalogs), this increases the number of INSs on our
  candidate list from 19\% of all INSs detected by the RASS/BSC in our
  survey region, to 31\%. Thus, we expect almost 1/3 of all INSs
  detected in the RASS/BSC in our survey region are in this selection.

\item In our previous work, we examined only RASS/BSC sources with a
  hardness ratio HR1$<$0, indicating an X-ray spectrum consistent
  with an effective temperature $kT_{\rm eff}<200\,{\rm
    eV}$\footnote{The text of R03, $\S3.1$ states this selection is
    HR1$>$0; this is a typographical error.  The selection which was
    applied, as discussed throughout the text, insured spectrally soft
    sources, with HR1$<$0.}.  In the present work, we include all
  RASS/BSC sources regardless of their spectrum.  This is due to the
  fact that, in the interim, we have discovered at least one INS with
  a spectrum harder than this limit \citep{rutledge08}, as well as the
  existence of the magnetars, a class of INS with $\kteff\sim
  600$\,eV. We therefore impose no spectral constraints on our
  RASS/BSC INS candidates.

\end{enumerate}

We also no longer impose the requirement that the \xray\ flux history
be consistent with no variability: there are now a handful of transient
magnetars \citep{ibrahim04, israel07, kumar08}, whose average flux
histories, absent giant flares and micro flares, still vary by 2 to 3
orders of magnitude -- brightening on timescales of $\sim$hours to
days, and subsequently fading over timescales of days to weeks (in
some cases, months).

The RASS/BSC spatial extendedness parameter, which characterizes X-ray 
source extendedness with a likelihood parameter ${\tt extl}=-log(P)$ 
where $P$ is the probability of the analysis concluding source extendedness 
from a point source, is not used as a selector in this work.  

A brief analysis has raised the possibility of false positives
for source extendedness in the RASS/BSC catalog.  There are, at
present, 105 RASS/BSC sources identified as isolated white dwarfs
(WDs) according to the SIMBAD\footnote{SIMBAD online listing,
  available at http://simbad.u-strasbg.fr/simbad/} database.  Of
these, twenty-one are listed with ${\tt extl}>10$, corresponding to a
chance detection from a non-extended source of 4\tee{-5}; we would
expect only 0.005 such detections, on average, in a population size of
105.

Of the twenty-one isolated white dwarfs with ${\tt extl} > 10$, one
has been observed with \chandra\ on-axis ($<$1\arcmin\ offset) and
without a grating: 1RXS~J103210.2+532941.  We find no evidence that
that WD is extended at \chandra\ resolution in the course of this
3\,ks observation.  Thus, we conclude that the {\tt extl} parameter
does not accurately represent the likelihood of spatial extent in the
0.1--2.4\,keV photon band for sources in the BSC.  (The alternative
scenario, that these white dwarfs are associated with extended
\xray\ emission that is variable on $\sim$year timescales, has not
otherwise been considered.)

Furthermore, there are two known observational classes of neutron
stars which are associated with supernova remnants (SNRs), the
compact-central objects (CCOs) and magnetars; since SNRs can be
spatially extended at \rosat/PSPC spatial resolution, discarding
candidate INSs on the basis of spatial extent, would miss these
classes of sources.  Given the systematic uncertainty in the accuracy
of the {\tt extl} parameter, we do not use spatial extent as a cut for
candidate INSs in the RASS/BSC.

Thus, we make no secondary cuts of INS candidates from the RASS/BSC
source selection other than the single selection criterion based on
\pnoid. We find 147 RASS/BSC sources with \pnoid$>$0.80, which are
listed in Table~\ref{tab:sources}.  So that we may compare results
with previous work (R03), we note that of these 147 RASS/BSC sources,
71 have hardness ratio HR1$<$0, implying a spectrum consistent with a
thermal spectrum of $kT_{\rm eff}<200$\,eV.


\subsection{Follow-up Swift Observations and X-ray Archival Data
  Analysis}

Following selection of the INS candidates, short ($\sim$ 1 \ksec)
follow-up observations with \swift/XRT \citep{xrt} were obtained on 92
of the candidates; these observations decrease the X-ray
positional uncertainty (the systematic positional error associated with 
\swift\ blind pointing observations is on the order of 3.5\arcsec
\citep{goad07}), and obtain (where possible) contemporaneous UV
observations with \swift/UVOT for counterpart identification with
off-band objects.

Assuming a thermal INS, an X-ray non-detection from a 1\ksec\ \swift/XRT
observation implies the RASS/BSC source has significantly decreased in
flux since the time of its detection in the all sky survey (1991/2).
For a nearby INS, one would expect negligible galactic absorption; for
a soft thermal source ($kT_{\rm eff}=100\,{\rm eV}$), at the limit of
the RASS/BSC (0.05 PSPC c/s), one expects 42 counts in a 1\,ks
integration; for $kT_{\rm eff}=500\,{\rm eV}$, one expects 39 counts.
For typical background count rates, in a 15\arcsec-radius region,
the average number of background counts is $\sim$0.26 \cpks,
making a detection above 5 counts secure with $>$4$\sigma$
confidence.  A non-detection ($<5$ counts) also implies a source from
which 40 counts was expected, has faded, with $>$6$\sigma$ confidence.
Thus, a non-detection during a \swift/XRT observation of a RASS/BSC
source implies the conclusion that the RASS/BSC source has
significantly decreased in X-ray flux, by a factor of $\approx$~8.

For flux conversions, where necessary, we assume an unabsorbed
power-law spectrum with $\alpha=2$.  Hydrogen column density was
obtained from the Leiden/Argentine/Bonn (LAB) Survey of Galactic HI
\citep{kalberla05}, using a 1\deg\ radius cone around each source
position.  An optical cross-identification catalog finds that the
spatial separation of optical sources within 75\arcsec\ of the
RASS/BSC position exhibits an excess above such in randomly chosen
(i.e. control) fields, detectable to a radial separation of 40\arcsec.
We therefore will consider possible counterparts which are offset from
the RASS/BSC position by $<$90\arcsec\ (to be conservative), or
3$\sigma$ statistical positional uncertainty of the RASS/BSC
localization, whichever is greater.  Positions given are in J2000
epoch.  Much of the X-ray analysis of \swift\ data was performed and
described in greater detail elsewhere \citep{shevchuk09}.

A number of RASS/BSC sources were redetected in the X-ray band 
with other instrumentation, and found to have decreased in flux.
To prevent source confusion with background AGN, we used the 
ROSAT/Deep survey \citep{hasinger98} to find the average number of 
expected AGN per square degree above the flux level of the redetected source.
If the probability that an as bright or brighter AGN
should lie as close or closer to the redetected source position 
was found to be $\leq1\%$, we consider the RASS/BSC source
and the redetected source to be associated.  

This calculation does not take into account the non-zero probability
of chance association with an unrelated X-ray source associated with
our own galaxy, such as a coronally active star or low mass X-ray
binary.  The effect of these sources on the probability of chance
association with the RASS/BSC source under consideration depends on the
unknown average surface density of galactic-associated X-ray sources
in the ROSAT passband, below the detection limit of the ROSAT
All-Sky-Survey, relative to the surface density as a function of flux
of AGN. If the surface density of galactic associated X-ray sources in
the direction of of the RASS/BSC source considered is below the
surface density used for AGN, then galactic-associated X-ray sources
will not significantly contribute to source confusion, and can be
neglected without altering our results here.  However, if it is above,
then galactic-associated X-ray sources would contribute to source
confusion, and could result in the mis-association of an unrelated
background X-ray source with a variable (and faded) ROSAT/BSC
source. To overcome this limitation requires an X-ray all-sky survey
with a deeper flux limit than the RASS, such as is planned to be
provided by eROSITA.  Here, we neglect the contribution of
galactic-associated X-ray sources in our calculation of source
confusion among X-ray sources fainter than the RASS/BSC flux limit.

The goal of these observations and data analyses is to determine for
each \xray\ source whether: (1) it is an INS; (2) it may be an INS
(``undetermined''); or (3) it can be excluded as an INS.


\subsection{Individual Objects}

Objects that have been observed and detected with \swift/XRT,
as well as nearby possible off-band counterparts, their 
probabilities of chance association, and our final classification
of each object as INS, not an INS and undetermined, 
are listed in Table~\ref{tab:swift}.

Objects that have been observed with \swift/XRT but were not
detected are listed in Table~\ref{tab:nonswift}.  Upper limits 
on fluxes were obtained by considering a non-detection to be
$<$5 counts for a $\sim$1 ks observations.  All of the objects
that were not detected with \swift have been classified as
undetermined.

If required, additional information about the analysis of specific 
objects from these tables is given in Appendix~\ref{sec:swiftobs}
Unless otherwise specified, all information and analyses from 
Appendix~\ref{sec:swiftobs} are obtained from \cite{shevchuk09}.

A full analysis of the remaining objects that were not observed 
with \swift, but rather with \chandra, \xmm, or \rosat/HRI, 
is given in Appendix~\ref{sec:otherobs}.


\section{Discussion}
\label{sec:discuss}

Here, we use the same calculation as previously (see $\S5$ of R03) to
derive upper-limits on the number of INSs detected in the full-sky
RASS/BSC.  We refer the reader to that reference for the description
of this calculation.  In short, with knowledge of the selection
efficiency of INSs detected in the RASS/BSC to our sample; the number
of sources in our sample which are known to be INSs; the number of
sources in our sample which are known to be not an INS; and the number
of sources in our sample which are of undetermined type (INS, or not),
one can place an upper-limit on the total number of INSs in the
RASS/BSC.

To estimate the the upper-limit on the number of INSs in our catalog,
we use the following values, where all variables take the same meaning
as in R03.  Of $A=150$ control sources, $B=46$ passed our
\pnoid\ selection; this implies that a real INS has probability $p$ 
of passing our INS selection, which can be represented as a 
binomial distribution:
\begin{equation}
\label{eq:bin}
P_{\rm XID}(p) = \frac{ p^{B}(1-p)^{A-B}} {\int_0^1 p^{B}(1-p)^{A-B}\;
dp}
\end{equation}

From the 15,205 RASS/BSC sources above our declination cut, a total
$T=147$ passed the \pnoid\ selection.  Based on the fraction of control
sources which passed our \pnoid\ selection (46/150=30.67\%), we expect
that 30.67\%, on average, of all INSs among the 15,205 RASS/BSC
sources are in our selected sample of $T=147$ RASS/BSC
sources.\footnote{For example, with no further calculations, this
would imply there are no more than 147/0.3067=480 INSs in the sample
of 15,205 RASS/BSC sources from which we selected.}

After examining the literature, \swift\ observations, and archival data
analysis, there are $N_{\rm INS, min}$=4 objects which we count as
INSs (1RXS J063354.1+174612 which is also known as Geminga;
1RXS~J130848.6+212708, 1RXS~J141256.0+792204 which is also known as
Calvera; and 1RXS~J160518.8+324907), 107 sources which are
not an INS, and $BG$=36 sources are found to be of ``undetermined'' 
type.  With these values, we can determine the un-nomalized probability
$P_{BG}(N)$ that $N$ sources in our selection of 147 are INSs:

\def\combfactor{
{\frac{   \left(\begin{array}{c} N \\ N_{\rm INS, min}  \end{array}\right)
\left( \begin{array}{c} T - N \\ T-BG-N_{\rm INS, min}  \end{array} \right) }
	{ \left( \begin{array}{c} T \\ T-BG  \end{array} \right) }}
}

\begin{equation}
\label{eq:comb}
P_{BG}(N)= {\combfactor}
\end{equation}
where $N_{\rm INS, min}\leq N \leq N_{\rm INS, max}$ and $N_{\rm INS, max}$
is the maximum number of INSs that could be present in our sample 
($N_{\rm INS, min}+BG$=40).  

Combining Eqs.~\ref{eq:bin} and \ref{eq:comb}
yields the unnormalized probability $P_{\rm INS}(M')$ that the total number 
of INSs in our survey field is $M'$: 

\begin{equation}
P_{\rm INS, un-normalized}(M')    = 
\sum_{N=N_{\rm INS, min}}^{{\rm min}(M', N_{\rm INS, max})}    \:
\left( \frac{P_{\rm BG}(N)} {\sum_{N=N_{\rm INS, min}}^{{\rm min}(M', N_{\rm INS, max})}      \:  P_{\rm BG}(N) }    \,    \int_0^1 P_{\rm XID}(p)\,  
\frac{p^N (1-p)^{M'-N} }{\int_0^1 p^N (1-p)^{M'-N}\; dp}\; dp\right)  
\end{equation}

Finally, to obtain the probability that there are $\geq M$ INSs in our
sample, we sum: 
\begin{equation}
\label{eq:pins}
P_{\rm INS}(\geq M)   = \sum^\infty_{M'=M} \frac{\: P_{\rm INS, un-normalized}(M')}
{\left(\sum^\infty_{M'=N_{\rm INS, min}} \:P_{\rm INS, un-normalized}(M') \right)}
\end{equation}

Using the above values, we place a 90\% confidence upper-limit on the
number of INSs among the 15,205 in our survey region to be $\leq$39;
and a 99\% confidence upper limit of $\leq$57. Rescaling
this to the total number of RASS/BSC sources full-sky, these
correspond to full-sky upper-limits on the number of INSs in the
RASS/BSC of $\leq$48 (90\% confidence) and $\leq$70 (99\% confidence).

{\em Variability}.  Of the 36 RASS/BSC sources in our selection which
we classify as ``undetermined'', 25 were undetected in a second-epoch
X-ray image (17\% of the total sample), which implies a significant
decrease in the X-ray flux.   While some classes of neutron stars
(e.g. magnetars) can exhibit significant fading, we can calculate
the number of INSs from classes which are not variable, by classifying
these 25 faded sources as ``not an INS'', and performing the
calculation again.  Doing so, we find the full-sky upper-limit on the
number of non-variable INSs in the RASS/BSC to be $\leq$39 (90\%
confidence) and $\leq$56 (99\% confidence).  

{\em Spectrally Soft, Non-Variable Sources}.  In previous work (R03),
only soft ($kT_{\rm eff}<200\, {\rm eV}$) sources which exhibit no
significant variability were considered as INSs.  In doing so,
full-sky upper-limits on the number of soft, non-variable INSs to be
$\leq$67 (90\% confidence) and $\leq$104 (99\% confidence) were
derived.  In the present analysis, we have 71 spectrally soft sources;
of these, 3 are identified as an INS, 66 are not an INS, and 2 are
undetermined. With these, we place new full-sky upper-limits on the
number of soft, non-variable INSs in the RASS/BSC of $\leq$31 (90\%
confidence) and $\leq$46 (99\%) confidence, both a little more than a
factor of two lower than the previously derived limits.

The above results are summarized in Table~\ref{tab:upperlimit}.


\subsection{A Targeted INS Candidate List}

We find \xray\ objects which were detected in the RASS/BSC, with
localizations $\sim$13\arcsec\ (1$\sigma$), which were redetected with
\swift/XRT follow-up observations and localized with $\sim$3.5\arcsec,
and which still have no likely counterparts in USNO-A2, 2MASS, NVSS,
IRAS, or in contemporaneous UVOT observations.

We are in the process of targeting these sources in \chandra\
observations, to bring the positional uncertainties down
$\leq$0.6\arcsec, and deep optical observations, to demonstrate
$F_X/F_{\rm opt}$ flux ratio $>$1000, which has been a robust
indicator of an INS.  We list these sources in Table~\ref{tab:ins}.


\subsection{The Relevant Populations, and eROSITA}

A next-generation \xray\ survey instrument, eROSITA, is currently
under construction and planned for flight as part of the Russian-led
Spectrum X-Gamma mission in 2011 \citep{erosita}.  eROSITA will reach
to $\times$100 deeper fluxes than the \rosat/All-Sky-Survey, and
provide 0.2--12\,keV energy coverage with a field-average point spread
function of 20\arcsec.  The full-sky survey is expected to achieve
0.5--2.0\,keV fluxes of 5.7\tee{-14}, approximately a factor of 30
below the RASS/BSC limit.  For a uniform disk population, this can be
expected to increase the number of detected INSs from the current
range of 8--50 to 240--1500 -- a significant increase, which would
permit a more detailed study of neutron star population models (see,
for example, \citet{posselt08} for a recent examination of population
synthesis prediction for cool INSs).


\section{Conclusions}
\label{sec:conclude}

We have placed upper-limits on the number of INSs -- indepdendent of
spectral and variability properties -- in the RASS/BSC, of $\leq$48
(90\% confidence) and $\leq$70 (99\% confidence).  When we limit
ourselves to spectrally soft, non-variable INSs, these limits are
$\leq$31 (90\% confidence) and $\leq$46 (99\% confidence) -- a factor
of 2 lower than limits derived previously (R03).

Using the same analysis as in previous work ($\S$ 5.2, R03), we can
place a limit on supernova rate of progenitors to these objects, using
a naive and optimistic model for the resulting sources.  
The following assumptions are made for this model:
\begin{enumerate}
 \item An INS production rate of $\gamma_{-2}$ per 100 yr
 \item INSs are produced in a flat disk with constant SNe 
rate per unit area, out to a radius of $R_{\rm disk}=15\,{\rm kpc}$
\item INSs are produced velocities of $<10\, {\rm kpc}\, {\rm Myr^{-1}}$
perpendicular to this disk
\item The resulting INSs maintain a
luminosity 2\tee{32} \cgslum for a period of $\tau=\ee{6}\, {\rm yr}$
in the \rosat/PSPC passband of 0.1-2.4 keV
\item The INSs exhibit an $\alpha=3$ spectrum
\item The limit on all INSs (assuming $\leq$50)
corresponds to a limit
\item The effects of absorption  (which are
likely to be important) are neglected (permitting INSs to be detected
to a distance of 10.3 kpc)
\end{enumerate}
We can then place a limit of
$\gamma_{-2}<0.019$ (90\% confidence), corresponding to an INS
birthrate of one per 5300 years. If we assume a spatial average
absorption of $N_H=3\tee{21}\, {\rm cm^{-2}}$ (limiting detection to
sources within a distance of 1.5 kpc), the limit is $\gamma_{\rm
  -2}<$0.8.

In the process of obtaining these limits, we have identified one new
confirmed INS \cite[Calvera]{rutledge08}.  After second-epoch
\xray\ observations with \swift, we find 9 INS candidates.  We are
presently following-up observationally on these candidates, with
better X-ray localizations and deep optical imaging, to place limits
of $F_X/F_{\rm opt}>1000$, on these sources.


\acknowledgements

The authors would like to express their appreciation to Neil Gehrels,
Jamie Kennea, and the \Swift\ team for carrying out the fill-in target
observations which form the basis for this paper.  This publication
makes use of data products from the Two Micron All Sky Survey, which
is a joint project of the University of Massachusetts and the Infrared
Processing and Analysis Center/California Institute of Technology,
funded by the National Aeronautics and Space Administration and the
National Science Foundation.  This research has made use of the
NASA/IPAC Infrared Science Archive, which is operated by the Jet
Propulsion Laboratory, California Institute of Technology, under
contract with the National Aeronautics and Space Administration.  The
Digitized Sky Surveys were produced at the Space Telescope Science
Institute under US Government grant NAGW-2166. The images of these
surveys are based on photographic data obtained using the Oschin
Schmidt Telescope on Palomar Mountain and the UK Schmidt
Telescope. The plates were processed into the present compressed
digital form with the permission of these institutions. The DPOSS
project was generously supported by the Norris Foundation.  RER and MT
are supported by the Discovery Grants Program of the Natural Sciences
and Engineering Research Council of Canada.

\bibliographystyle{apj_8}
\bibliography{complete,ins2}

\begin{thebibliography}{65}
\expandafter\ifx\csname natexlab\endcsname\relax\def\natexlab#1{#1}\fi

\bibitem[{{Arras} {et~al.}(2004){Arras}, {Cumming}, \& {Thompson}}]{arras04}
{Arras}, P., {Cumming}, A., \& {Thompson}, C. 2004, \apjl, 608, L49

\bibitem[{{Arzoumanian} {et~al.}(2002){Arzoumanian}, {Chernoff}, \&
  {Cordes}}]{arzoumanian02}
{Arzoumanian}, Z., {Chernoff}, D.~F., \& {Cordes}, J.~M. 2002, \apj, 568, 289

\bibitem[{{Beuermann} {et~al.}(1991){Beuermann}, {Thomas}, \&
  {Pietsch}}]{beuermann91}
{Beuermann}, K., {Thomas}, H.-C., \& {Pietsch}, W. 1991, \aap, 246, L36

\bibitem[{{Blaes} \& {Madau}(1993)}]{blaes93}
{Blaes}, O. \& {Madau}, P. 1993, \apj, 403, 690

\bibitem[{{B{\"o}hringer} {et~al.}(2004){B{\"o}hringer}, {Schuecker}, {Guzzo},
  {Collins}, {Voges}, {Cruddace}, {Ortiz-Gil}, {Chincarini}, {De Grandi},
  {Edge}, {MacGillivray}, {Neumann}, {Schindler}, \& {Shaver}}]{bohringer04}
{B{\"o}hringer}, H. {et al.}\  2004, \aap, 425, 367

\bibitem[{{Burrows} {et~al.}(2003){Burrows}, {Hill}, {Nousek}, {Wells},
  {Short}, {Ambrosi}, {Chincarini}, {Citterio}, \& {Tagliaferri}}]{xrt}
{Burrows}, D.~N. {et al.}\  2003, in X-Ray and Gamma-Ray Telescopes and
  Instruments for Astronomy. Edited by Joachim E. Truemper, Harvey D.
  Tananbaum. Proceedings of the SPIE, Volume 4851, pp. 1320-1325 (2003)., 1320

\bibitem[{{Camilo} {et~al.}(2000){Camilo}, {Kaspi}, {Lyne}, {Manchester},
  {Bell}, {D'Amico}, {McKay}, \& {Crawford}}]{camilo00}
{Camilo}, F., {Kaspi}, V.~M., {Lyne}, A.~G., {Manchester}, R.~N., {Bell},
  J.~F., {D'Amico}, N., {McKay}, N. P.~F., \& {Crawford}, F. 2000, \apj, 541,
  367

\bibitem[{{Camilo} {et~al.}(1994){Camilo}, {Thorsett}, \&
  {Kulkarni}}]{camilo94}
{Camilo}, F., {Thorsett}, S.~E., \& {Kulkarni}, S.~R. 1994, \apjl, 421, L15

\bibitem[{{Cordes} \& {Chernoff}(1998)}]{cordes98}
{Cordes}, J.~M. \& {Chernoff}, D.~F. 1998, \apj, 505, 315

\bibitem[{{Dotani} {et~al.}(1999){Dotani}, {Asai}, \& {Greiner}}]{dotani99}
{Dotani}, T., {Asai}, K., \& {Greiner}, J. 1999, \pasj, 51, 519

\bibitem[{{Ebeling} {et~al.}(2007){Ebeling}, {Barrett}, {Donovan}, {Ma},
  {Edge}, \& {van Speybroeck}}]{ebeling07}
{Ebeling}, H., {Barrett}, E., {Donovan}, D., {Ma}, C.-J., {Edge}, A.~C., \&
  {van Speybroeck}, L. 2007, \apjl, 661, L33

\bibitem[{{Faucher-Gigu{\`e}re} \& {Kaspi}(2006)}]{faucher06}
{Faucher-Gigu{\`e}re}, C.-A. \& {Kaspi}, V.~M. 2006, \apj, 643, 332

\bibitem[{{Fleming} {et~al.}(1996){Fleming}, {Snowden}, {Pfeffermann}, {Briel},
  \& {Greiner}}]{fleming96}
{Fleming}, T.~A., {Snowden}, S.~L., {Pfeffermann}, E., {Briel}, U., \&
  {Greiner}, J. 1996, \aap, 316, 147

\bibitem[{{Fuhrmeister} \& {Schmitt}(2003)}]{fuhrmeister03}
{Fuhrmeister}, B. \& {Schmitt}, J.~H.~M.~M. 2003, \aap, 403, 247

\bibitem[{{Furusho} {et~al.}(2003){Furusho}, {Yamasaki}, \&
  {Ohashi}}]{furusho03}
{Furusho}, T., {Yamasaki}, N.~Y., \& {Ohashi}, T. 2003, \apj, 596, 181

\bibitem[{{Goad} {et~al.}(2007){Goad}, {Tyler}, {Beardmore}, {Evans}, {Rosen},
  {Osborne}, {Starling}, {Marshall}, {Yershov}, {Burrows}, {Gehrels}, {Roming},
  {Moretti}, {Capalbi}, {Hill}, {Kennea}, {Koch}, \& {vanden Berk}}]{goad07}
{Goad}, M.~R. {et al.}\  2007, \aap, 476, 1401

\bibitem[{{Gonthier} {et~al.}(2004){Gonthier}, {Van Guilder}, \&
  {Harding}}]{gonthier04}
{Gonthier}, P.~L., {Van Guilder}, R., \& {Harding}, A.~K. 2004, \apj, 604, 775

\bibitem[{{Haakonsen} \& {Rutledge}(2009)}]{haakonsen09}
{Haakonsen}, C. \& {Rutledge}, R. 2009, \apjs, submitted

\bibitem[{{Haberl} {et~al.}(1994){Haberl}, {Throstensen}, {Motch},
  {Schwarzenberg-Czerny}, {Pakull}, {Shambrook}, \& {Pietsch}}]{haberl94}
{Haberl}, F., {Throstensen}, J.~R., {Motch}, C., {Schwarzenberg-Czerny}, A.,
  {Pakull}, M., {Shambrook}, A., \& {Pietsch}, W. 1994, \aap, 291, 171

\bibitem[{{Halpern}(1992)}]{halpern92}
{Halpern}, J.~P. 1992, in EUVE Proposal, 38--+

\bibitem[{{Hambaryan} {et~al.}(2002){Hambaryan}, {Hasinger}, {Schwope}, \&
  {Schulz}}]{hambaryan02}
{Hambaryan}, V., {Hasinger}, G., {Schwope}, A.~D., \& {Schulz}, N.~S. 2002,
  \aap, 381, 98

\bibitem[{{Hasinger} {et~al.}(1998){Hasinger}, {Burg}, {Giacconi}, {Schmidt},
  {Truemper}, \& {Zamorani}}]{hasinger98}
{Hasinger}, G., {Burg}, R., {Giacconi}, R., {Schmidt}, M., {Truemper}, J., \&
  {Zamorani}, G. 1998, \aap, 329, 482

\bibitem[{{Ibrahim} {et~al.}(2004){Ibrahim}, {Markwardt}, {Swank}, {Ransom},
  {Roberts}, {Kaspi}, {Woods}, {Safi-Harb}, {Balman}, {Parke}, {Kouveliotou},
  {Hurley}, \& {Cline}}]{ibrahim04}
{Ibrahim}, A.~I. {et al.}\  2004, \apjl, 609, L21

\bibitem[{{Imanishi} {et~al.}(2001){Imanishi}, {Koyama}, \&
  {Tsuboi}}]{imanishi01}
{Imanishi}, K., {Koyama}, K., \& {Tsuboi}, Y. 2001, \apj, 557, 747

\bibitem[{{Israel} {et~al.}(2007){Israel}, {Campana}, {Dall'Osso}, {Muno},
  {Cummings}, {Perna}, \& {Stella}}]{israel07}
{Israel}, G.~L., {Campana}, S., {Dall'Osso}, S., {Muno}, M.~P., {Cummings}, J.,
  {Perna}, R., \& {Stella}, L. 2007, \apj, 664, 448

\bibitem[{{Kalberla} {et~al.}(2005){Kalberla}, {Burton}, {Hartmann}, {Arnal},
  {Bajaja}, {Morras}, \& {P{\"o}ppel}}]{kalberla05}
{Kalberla}, P.~M.~W., {Burton}, W.~B., {Hartmann}, D., {Arnal}, E.~M.,
  {Bajaja}, E., {Morras}, R., \& {P{\"o}ppel}, W.~G.~L. 2005, \aap, 440, 775

\bibitem[{{Kaplan} {et~al.}(2002){Kaplan}, {van Kerkwijk}, \&
  {Anderson}}]{kaplan02b}
{Kaplan}, D.~L., {van Kerkwijk}, M.~H., \& {Anderson}, J. 2002, \apj, 571, 447

\bibitem[{{Kaplan} {et~al.}(2007){Kaplan}, {van Kerkwijk}, \&
  {Anderson}}]{kaplan07}
--- 2007, ArXiv Astrophysics e-prints, astro-ph/0703343

\bibitem[{{Kong} {et~al.}(2006){Kong}, {Charles}, {Homer}, {Kuulkers}, \&
  {O'Donoghue}}]{kong06}
{Kong}, A.~K.~H., {Charles}, P.~A., {Homer}, L., {Kuulkers}, E., \&
  {O'Donoghue}, D. 2006, \mnras, 368, 781

\bibitem[{{Kumar} \& {Safi-Harb}(2008)}]{kumar08}
{Kumar}, H.~S. \& {Safi-Harb}, S. 2008, \apjl, 678, L43

\bibitem[{{Lira} {et~al.}(2000){Lira}, {Lawrence}, \& {Johnson}}]{lira00}
{Lira}, P., {Lawrence}, A., \& {Johnson}, R.~A. 2000, \mnras, 319, 17

\bibitem[{{Liu} {et~al.}(2001){Liu}, {van Paradijs}, \& {van den
  Heuvel}}]{liu01}
{Liu}, Q.~Z., {van Paradijs}, J., \& {van den Heuvel}, E.~P.~J. 2001, \aap,
  368, 1021

\bibitem[{{Liu} {et~al.}(2006){Liu}, {van Paradijs}, \& {van den
  Heuvel}}]{liu06}
--- 2006, \aap, 455, 1165

\bibitem[{{Liu} {et~al.}(2007){Liu}, {van Paradijs}, \& {van den
  Heuvel}}]{liu07}
--- 2007, \aap, 469, 807

\bibitem[{{Manchester} {et~al.}(2005){Manchester}, {Hobbs}, {Teoh}, \&
  {Hobbs}}]{manchester05}
{Manchester}, R.~N., {Hobbs}, G.~B., {Teoh}, A., \& {Hobbs}, M. 2005, \aj, 129,
  1993

\bibitem[{{Marino} {et~al.}(2000){Marino}, {Micela}, \& {Peres}}]{marino00}
{Marino}, A., {Micela}, G., \& {Peres}, G. 2000, \aap, 353, 177

\bibitem[{{Motch} {et~al.}(1999){Motch}, {Haberl}, {Zickgraf}, {Hasinger}, \&
  {Schwope}}]{motch99}
{Motch}, C., {Haberl}, F., {Zickgraf}, F.-J., {Hasinger}, G., \& {Schwope},
  A.~D. 1999, \aap, 351, 177

\bibitem[{{Page} {et~al.}(2007){Page}, {Geppert}, \& {K{\"u}ker}}]{page07}
{Page}, D., {Geppert}, U., \& {K{\"u}ker}, M. 2007, \apss, 308, 403

\bibitem[{{Page} {et~al.}(2004){Page}, {Lattimer}, {Prakash}, \&
  {Steiner}}]{page04}
{Page}, D., {Lattimer}, J.~M., {Prakash}, M., \& {Steiner}, A.~W. 2004, \apjs,
  155, 623

\bibitem[{{Panzera} {et~al.}(2003){Panzera}, {Campana}, {Covino}, {Lazzati},
  {Mignani}, {Moretti}, \& {Tagliaferri}}]{panzera03}
{Panzera}, M.~R., {Campana}, S., {Covino}, S., {Lazzati}, D., {Mignani}, R.~P.,
  {Moretti}, A., \& {Tagliaferri}, G. 2003, \aap, 399, 351

\bibitem[{{Perna} {et~al.}(2003){Perna}, {Narayan}, {Rybicki}, {Stella}, \&
  {Treves}}]{perna03}
{Perna}, R., {Narayan}, R., {Rybicki}, G., {Stella}, L., \& {Treves}, A. 2003,
  \apj, 594, 936

\bibitem[{{Perryman} {et~al.}(1997){Perryman}, {Lindegren}, {Kovalevsky},
  {Hoeg}, {Bastian}, {Bernacca}, {Cr{\'e}z{\'e}}, {Donati}, {Grenon}, {van
  Leeuwen}, {van der Marel}, {Mignard}, {Murray}, {Le Poole}, {Schrijver},
  {Turon}, {Arenou}, {Froeschl{\'e}}, \& {Petersen}}]{perryman97}
{Perryman}, M.~A.~C. {et al.}\  1997, \aap, 323, L49

\bibitem[{{Popov} {et~al.}(2006{\natexlab{a}}){Popov}, {Grigorian}, {Turolla},
  \& {Blaschke}}]{popov06b}
{Popov}, S., {Grigorian}, H., {Turolla}, R., \& {Blaschke}, D.
  2006{\natexlab{a}}, \aap, 448, 327

\bibitem[{{Popov}(2001)}]{popov01}
{Popov}, S.~B. 2001, in "Astrophysical Sources of High Energy Particles \&
  Radiation", Erice, Italy November 2000., astro-ph/0101031

\bibitem[{{Popov} {et~al.}(2000){Popov}, {Colpi}, {Prokhorov}, {Treves}, \&
  {Turolla}}]{popov00b}
{Popov}, S.~B., {Colpi}, M., {Prokhorov}, M.~E., {Treves}, A., \& {Turolla}, R.
  2000, \apjl, 544, L53

\bibitem[{{Popov} {et~al.}(2003){Popov}, {Colpi}, {Prokhorov}, {Treves}, \&
  {Turolla}}]{popov03}
--- 2003, \aap, 406, 111

\bibitem[{{Popov} {et~al.}(2006{\natexlab{b}}){Popov}, {Grigorian}, \&
  {Blaschke}}]{popov06}
{Popov}, S.~B., {Grigorian}, H., \& {Blaschke}, D. 2006{\natexlab{b}}, \prc,
  74, 025803

\bibitem[{{Posselt} {et~al.}(2007){Posselt}, {Popov}, {Haberl}, {Tr{\"u}mper},
  {Turolla}, \& {Neuh{\"a}user}}]{posselt07}
{Posselt}, B., {Popov}, S.~B., {Haberl}, F., {Tr{\"u}mper}, J., {Turolla}, R.,
  \& {Neuh{\"a}user}, R. 2007, \apss, 308, 171

\bibitem[{{Posselt} {et~al.}(2008){Posselt}, {Popov}, {Haberl}, {Tr{\"u}mper},
  {Turolla}, \& {Neuh{\"a}user}}]{posselt08}
--- 2008, \aap, 482, 617

\bibitem[{{Predehl} {et~al.}(2007){Predehl}, {Andritschke}, {Bornemann},
  {Br{\"a}uninger}, {Briel}, {Brunner}, {Burkert}, {Dennerl}, {Eder},
  {Freyberg}, {Friedrich}, {F{\"u}rmetz}, {Hartmann}, {Hartner}, {Hasinger},
  {Herrmann}, {Holl}, {Huber}, {Kendziorra}, {Kink}, {Meidinger}, {M{\"u}ller},
  {Pavlinsky}, {Pfeffermann}, {Roh{\'e}}, {Santangelo}, {Schmitt}, {Schwope},
  {Steinmetz}, {Str{\"u}der}, {Sunyaev}, {Tiedemann}, {Vongehr}, {Wilms},
  {Erhard}, {Gutruf}, {Jugler}, {Kampf}, {Graue}, {Citterio}, {Valsecci},
  {Vernani}, \& {Zimmerman}}]{erosita}
{Predehl}, P. {et al.}\  2007, in Society of Photo-Optical Instrumentation
  Engineers (SPIE) Conference Series, Vol. 6686, Society of Photo-Optical
  Instrumentation Engineers (SPIE) Conference Series

\bibitem[{{Ramsay} {et~al.}(2000){Ramsay}, {Cropper}, {Wu}, {Mason}, \&
  {Hakala}}]{ramsay00}
{Ramsay}, G., {Cropper}, M., {Wu}, K., {Mason}, K.~O., \& {Hakala}, P. 2000,
  \mnras, 311, 75

\bibitem[{{Read} {et~al.}(1997){Read}, {Ponman}, \& {Strickland}}]{read97}
{Read}, A.~M., {Ponman}, T.~J., \& {Strickland}, D.~K. 1997, \mnras, 286, 626

\bibitem[{{Rutledge} {et~al.}(2000){Rutledge}, {Bildsten}, {Brown}, {Pavlov},
  \& {Zavlin}}]{rutledge00}
{Rutledge}, R.~E., {Bildsten}, L., {Brown}, E.~F., {Pavlov}, G.~G., \&
  {Zavlin}, V.~E. 2000, \apj, 529, 985

\bibitem[{{Rutledge} {et~al.}(2008){Rutledge}, {Fox}, \&
  {Shevchuk}}]{rutledge08}
{Rutledge}, R.~E., {Fox}, D.~B., \& {Shevchuk}, A.~H. 2008, \apj, 672, 1137

\bibitem[{{Rutledge} {et~al.}(2003){Rutledge}, {Fox}, {Bogosavljevic}, \&
  {Mahabal}}]{rutledge03}
{Rutledge}, R.~E., {Fox}, D.~W., {Bogosavljevic}, M., \& {Mahabal}, A. 2003,
  \apj, 598, 458

\bibitem[{{Schwarz} {et~al.}(2005){Schwarz}, {Reinsch}, {Beuermann}, \&
  {Burwitz}}]{schwarz05}
{Schwarz}, R., {Reinsch}, K., {Beuermann}, K., \& {Burwitz}, V. 2005, \aap,
  442, 271

\bibitem[{{Shevchuk} {et~al.}(2009){Shevchuk}, Fox, Letcavage, Turner, \&
  Rutledge}]{shevchuk09}
{Shevchuk}, A., Fox, F.~B., Letcavage, R.~J., Turner, M., \& Rutledge, R.~E.
  2009, \apj, in progress

\bibitem[{{Silber} {et~al.}(1994){Silber}, {Vrtilek}, \& {Raymond}}]{silber94}
{Silber}, A., {Vrtilek}, S.~D., \& {Raymond}, J.~C. 1994, \apj, 425, 829

\bibitem[{{Singh} {et~al.}(1995){Singh}, {Barrett}, {Schlegel}, {White},
  {Szkody}, {Silber}, {Hoard}, \& {Fierce}}]{singh95a}
{Singh}, K.~P., {Barrett}, P., {Schlegel}, E., {White}, N.~E., {Szkody}, P.,
  {Silber}, A., {Hoard}, D.~W., \& {Fierce}, E. 1995, \iaucirc, 6195, 2

\bibitem[{{Timmes} {et~al.}(1996){Timmes}, {Woosley}, \& {Weaver}}]{timmes96}
{Timmes}, F.~X., {Woosley}, S.~E., \& {Weaver}, T.~A. 1996, \apj, 457, 834

\bibitem[{{Treves} {et~al.}(2000){Treves}, {Turolla}, {Zane}, \&
  {Colpi}}]{treves00}
{Treves}, A., {Turolla}, R., {Zane}, S., \& {Colpi}, M. 2000, \pasp, 112, 297

\bibitem[{{Voges} {et~al.}(1999){Voges}, {Aschenbach}, {Boller},
  {Br\"auninger}, {Briel}, {Burkert}, {Dennerl}, {Englhauser}, {Gruber},
  {Haberl}, {Hartner}, {Hasinger}, {K\"urster}, {Pfeffermann}, {Pietsch},
  {Predehl}, {Rosso}, {Schmitt}, {Truemper}, \& {Zimmermann}}]{voges99}
{Voges}, W. {et al.}\  1999, \aap, 349, 389

\bibitem[{{Watson} {et~al.}(1987){Watson}, {King}, \& {Williams}}]{watson87}
{Watson}, M.~G., {King}, A.~R., \& {Williams}, G.~A. 1987, \mnras, 226, 867

\bibitem[{{Wijnands} {et~al.}(2002){Wijnands}, {Heinke}, \&
  {Grindlay}}]{wijnands02}
{Wijnands}, R., {Heinke}, C.~O., \& {Grindlay}, J.~E. 2002, \apj, 572, 1002

\bibitem[{{Wilson} {et~al.}(2003){Wilson}, {Patel}, {Kouveliotou}, {Jonker},
  {van der Klis}, {Lewin}, {Belloni}, \& {M{\'e}ndez}}]{wilson03}
{Wilson}, C.~A., {Patel}, S.~K., {Kouveliotou}, C., {Jonker}, P.~G., {van der
  Klis}, M., {Lewin}, W.~H.~G., {Belloni}, T., \& {M{\'e}ndez}, M. 2003, \apj,
  596, 1220

\end{thebibliography}


\clearpage

\appendix

\section{Objects Observed with \swift/XRT} \label{sec:swiftobs}

{ \em 1RXS J012428.1-335504. } - We note the position of the 2MASS source is inconsistent with
the position of the UVOT source (offset by $1.3\pm0.4$\arcsec); this
is consistent with the classification for this  in SIMBAD
as a high proper-motion star. \done
 
{ \em 1RXS J014205.0+213045. } - 
Examination of the DSS R-band images centered on the XRT position 
finds a group of $\sim$3 point sources with a surrounding nebulosity. 
The XRT source has been found to be extended, with probability that
the observed X-ray brightness is consistent with the PSF of \swift/XRT
 of \ee{-20.3}.  \notanins. \done 
 
{ \em 1RXS J015311.6-210545. } - 
 We note the position of the 2MASS source is inconsistent with
the position of the UVOT source (offset by $2.1\pm0.3$\arcsec); this
is consistent with the classification for this  in SIMBAD
as a high proper-motion star. 
 
{ \em 1RXS J020146.5+011717. } - Examination of the DSS and 2MASS images of the
counterpart reveals a possible optical/IR binary.

{ \em 1RXS J024946.0-382540. } - The XRT source may also be associated with 
2MASS~024945.8$-$382440.1 which may, itself, be a binary counterpart to
 2MASS~J024945.9-382536.6. 

{ \em 1RXS J031413.7-223533. } - This source has previously been
classified as a nova/star  \citep{watson87,beuermann91}, and
classified as not an INS (R03). 
 
{ \em 1RXS J040314.6-360927. } -  We note that the
position of the 2MASS source is inconsistent with the position of the
UVOT source (offset by $2.0\pm0.3$\arcsec).
 
{ \em 1RXS J040913.8+110833. } - 
The XRT source has been found to be extended, with probability that
the observed X-ray brightness is consistent with the PSF of \swift/XRT
of \ee{-43.6}.  

{ \em 1RXS J041215.8+644407. } -  
We note that the 2MASS and UVOT source, however, are significantly offset
from one another (5.7\arcsec\ppm0.2\arcsec). 

{ \em 1RXS J043334.8+204437. } - 
 We note the position of the 2MASS source is inconsistent with the
position of the UVOT source (offset by $3.9\pm0.3$\arcsec).
 
{ \em 1RXS J044048.0+292440. } - No source is detected in the UVOT
image (M2$>$21.5, 3$\sigma$). 
The XRT source has been found to be only marginally consistent 
with a point source, with probability that
the observed X-ray brightness is consistent with the PSF of \swift/XRT 
of \ee{-2.6 }.  \unde.

{ \em 1RXS J050909.9+152740. } -
 We note the position of the 2MASS source is inconsistent with
the position of the UVOT source (offset by $3.5\pm0.2$\arcsec).

{ \em 1RXS J051315.9+025227. } - \swift/XRT observation on 2006 March 26 18:57 UT
(\obsid=00035532001)
detects an apparently saturated X-ray source in the field, for which
the standard astrometry is unreliable.  Nonetheless, the best fit of
this astrometry produces a position which is 
 approximately 1\arcmin\ from the RASS/BSC position (the next nearest other
RASS/BSC source is more than 15\arcmin\ away), but with an
unquantifiable systematic uncertainty.  The count rate appeared to be
sufficiently great ($>1$ count \persec) that the on-board rejection
removed most of the X-ray counts from the datastream, which is in
excess of the count rate which would be expected from the RASS/BSC
count rate (0.09 counts \persec).  
An XRT observation on 2006 March 29 04:52 UT (\obsid=00035532002) re-detects 
the X-ray source, at $05\h13\m15\s.9,+02\deg51\arcmin43\arcsec$.  
There is an optical source USNO~051317.4+025140.5 (B=5.97, R=3.81), 
23\arcsec\ away from the XRT source. SV*~ZI~359 is located at 
05\h13\m17\s.4,+02\deg51\arcmin40\arcsec\ \citep{perryman97}.   
SV*~ZI~359 is classified on SIMBAD as a  K0.5III star, and a spectroscopic
binary.  At the X-ray flux from the RASS/BSC, the
implied X-ray luminosity at the distance of SV* ZI 359 (105\ppm8 pc,
\cite{perryman97}) is $L_X$=1.2\tee{30} \cgslum,
which is consistent with the luminosity expected from the surface of
such a star.
 
{ \em 1RXS J051354.0+023722. } - We note that the
XRT source is offset from the RASS/BSC position by 50\arcsec\, and 
another XRT source with a lower count rate is located 160\arcsec\ 
from the RASS/BSC position.

{ \em 1RXS J051723.3-352152. } - 
 We note the position of the 2MASS source is inconsistent with
the position of the UVOT source (offset by $1.9\pm0.3$\arcsec); this
is consistent with the classification for this  in SIMBAD
as a high proper-motion star. 

{ \em 1RXS J055734.5-273534. } - This source has been observed with 
\swift/XRT three times: on 2006 April 14 02:04:28 UT (\obsid=00035513001, 
duration 769 s); on 2006 April 20 23:14:06 UT (\obsid=00035513002, duration 
1.1 \ksec) and on 2006 April 27 UT 06:22:55 (\obsid=00035513003, duration 
1.3 \ksec).  For all three observations, no X-ray source was detected 
in the XRT field within 90\arcsec\ of the RASS/BSC position \citep{shevchuk09}.
Since $<$5 counts were detected during each observation, 
the upper limit on the flux is 2.1\tee{-13} \cgsflux (0.1-2.4 keV). 
 
{ \em 1RXS J075556.7+832310. } -  We note the position of 
the 2MASS source is inconsistent with
the position of the UVOT source (offset by $4.1\pm0.2$\arcsec).
 
{ \em 1RXS J094454.2-122047. } -  We note the position of the 
2MASS source is inconsistent with the
position of the UVOT source (offset by $2.6\pm0.2$\arcsec); this is
consistent with the classification for this object in SIMBAD as a high
proper-motion star.

{ \em 1RXS J120711.0+364745. } -   We note the position of the 
2MASS source is inconsistent with the position of the
UVOT source (offset by $1.0\pm0.3$\arcsec).
 
{ \em 1RXS J121900.7+110727. } - We note the position of the 
2MASS source is inconsistent with the
position of the UVOT source (offset by $7.5\pm0.3$\arcsec).

{ \em 1RXS J122940.6+181645. } - This object was previously 
observed and undetected (R03).

{ \em 1RXS J125947.9+275636. } - We
note the position of the 2MASS source is inconsistent with the
position of the UVOT source (offset by $2.5\pm0$\arcsec).  The XRT
X-ray source appears extended to visual inspection, with an
approximate width of 5\arcmin\ in diameter.  This is consistent with the
position for the associated RASS/BSC source.  We conclude the XRT
object is the RASS/BSC source, which corresponds to the Coma Cluster
of galaxies.

{ \em 1RXS J131011.9+474521. } - We note the 2MASS and UVOT sources are offset by
7.6\ppm0.4\arcsec, indicating a high proper motion star, which is also
apparent by comparing DSS B-band images from 1955 March 18 and
1995 March 27. 

{ \em 1RXS J132041.2-030010. } - This source has been observed with 
\swift/XRT twice: on 2006 June 10 00:05:06 UT (\obsid=00035555001, duration
287 s); and on 2007 January 03 04:00:30 UT (\obsid=00035555003, duration 1.3 \ksec).
For both observations, no X-ray source was detected 
in the XRT field within 90\arcsec\ of the RASS/BSC position.

{ \em 1RXS J133825.0-251634. } - We note the position of the 
2MASS source is inconsistent with the
position of the UVOT source (offset by $2.1\pm0.3$\arcsec).

{ \em 1RXS J141703.1+314249. } - Comparing archival DSS
images (1950 April 12) with more recent 2MASS images show the
identified 2MASS counterpart has exhibited significant proper motion
over this epoch.
 
{ \em 1RXS J142644.1+500633. } - To place the flux
upper-limit, we note detection of a marginal (3$\sigma$) source in the
field, with a count rate of 25.7 \cpks, which corresponds to a
flux of 7.4\tee{-13} \cgsflux (0.5-10 keV), which we adopt as a
conservative flux upper-limit. 

{ \em 1RXS J143652.6+582104. } -  We note the position of the 
2MASS source is inconsistent with the
position of the UVOT source (offset by $6.4\pm0.4$\arcsec).  A
comparison of archival DSS images (1955 March 16) indicates the 2MASS
source (observed 2000 March 3) exhibits significant proper motion over
this epoch. 
  
{ \em 1RXS J153840.1+592118. } -  We note the position of the 
2MASS source is inconsistent with the position of the
UVOT source (offset by $1.4\pm0.4$\arcsec).

{ \em 1RXS J164020.0+673612. } - We note the position of the 
2MASS source is inconsistent with the
position of the UVOT source (offset by $3.3\pm0.3$\arcsec).

{ \em 1RXS J171502.4-333344. } -  
Due to the large number of 2MASS point sources within the RASS/BSC error circle
(11 2MASS sources within 16\arcsec\ radius or $2\sigma$),
the probability that the RASS/BSC X-ray source is not associated with any 
2MASS source is \ee{-2.2}.  The most likely association was found to be with
2MASS~J171502.19-333339.8 ($J=7.9$), with probability of association
 $P_{\rm id}$=0.782 \citep{haakonsen09}. 
The XRT source has been found to be only marginally 
consistent with a point source, with probability that
the observed X-ray brightness is consistent with the PSF of \swift/XRT
of \ee{-2.0} \citep{shevchuk09}.  

{ \em 1RXS J205549.4+435216. } -  We note that the
2MASS colors for this bright IR source would be unusual for a low-mass
star (J=6.36, H=5.51, K=5.04), and the object warrants IR spectral
analysis.  \notanins\ \citep{shevchuk09}.  \done
 
 { \em 1RXS J212700.3+101108. } - The XRT source has been found to be 
consistent with a point source, with probability that
the observed X-ray brightness is consistent with the PSF of \swift/XRT 
of 0.40.
 
{ \em 1RXS J213944.3+595016. } - The XRT source has been found to be 
consistent with a point source,  with probability that
the observed X-ray brightness is consistent with the PSF of \swift/XRT
of 0.32

{ \em 1RXS J230340.4-352420. } -  This source has been observed with 
\swift/XRT three times: on 2006 January 05 12:04:41 UT (\obsid=00035579001, 
duration 240 s); on 2006 April 19 00:32:11 UT (\obsid=00035579002, 
duration 1.3 \ksec); and on 2006 May 01 03:13:03 TT (\obsid=00035579003,
duration 3.2 \ksec).  For all three observations, no X-ray source was detected 
in the XRT field within 90\arcsec\ of the RASS/BSC position.

{ \em 1RXS J231543.7-122159. } -  We note the position of the 2MASS 
source is inconsistent with the position of the UVOT source (offset 
by $4.3\pm0.4$\arcsec); this is consistent with the classification 
for this in SIMBAD as a high proper-motion star. 

{ \em 1RXS J231728.9+193651. } - We note the position of the 
2MASS source is inconsistent with the
position of the UVOT source (offset by $3.4\pm0.2$\arcsec).

{ \em 1RXS J234421.2+213601. } -   We note the position of the 
2MASS source is inconsistent with the
position of the UVOT source (offset by $4.0\pm0.3$\arcsec).

{ \em 1RXS J234836.5-273935. } -  We note the position of the 
2MASS source is inconsistent with the
position of the UVOT source (offset by $4.9\pm0.4$\arcsec).

\section{Objects Observed with Other Instrumentation} \label{sec:otherobs}

{ \em 1RXS J001832.0+162634. } - This object has been identified as the
galaxy cluster ClG 0015.9+1609, which has been observed with
\chandra\ and found to be significantly extended
\citep{ebeling07}.  \notanins.  \done\
  
{ \em 1RXS J004330.9+411452. } - \chandra/ACIS-I observation of M31 (\obsid=1585, 
start time 2001 November 19 18:25:51 UT, duration 4.88 \ksec) was examined.  No sources
were found by {\tt wavdetect}  within 70\arcsec\ of the RASS/BSC position.  
The \chandra\ observation only reveals the highly resolved M31
galaxy, with no point source in the vicinity of the RASS/BSC position.

We note the presence of an X-ray source CXO~J004337.27+411443.4
approximately 75\arcsec\ from the RASS/BSC position, with a count rate
of 45.7\ppm3.7~\cpks, which corresponds to a flux of 
6.3\tee{-13}\cgsflux(0.1-2.4 keV). The total
positional uncertainty of the RASS/BSC source is $\ppm$26\arcsec, placing
this object $2.9\sigma$ from the RASS/BSC position.  This \chandra\
source is at the same position as an {\em XMM} detected X-ray source
XMMM31~J004337.2+411444, which is associated with a globular cluster
at the distance of M31; this object has, at the 770 kpc distance of
M31, a luminosity of 1.7\tee{37}\cgslum.  The observed modestly lower
($\times 3.2$) count rate is consistent with the spectral uncertainty
and possible intrinsic source variability which may be associated with
a low-mass X-ray binary in a globular cluster. 
The probability of chance association between the \chandra/ACIS-I source
and a background AGN is \pchance=\ee{-2.3}.

We therefore conclude that
1RXS~J4330.9+411452 is the X-ray source CXO~J004337.27+411443.4, and
the X-ray source XMMM31~J004337.2+41144. \notanins.  \done

{ \em 1RXS J004704.8-204743. } - The galaxy NGC~247 was observed with
\rosat/HRI \citep{lira00}.  Two bright X-ray sources were found south
of the galaxy (denoted X1 and X2 in \cite{lira00}).  By visual
inspection, the RASS/BSC source appears spatially coincident with X1
(and inconsistent with X2, which lies $\sim$1\arcmin\ away).  X1 is
found to have a very soft spectral distribution \citep{read97}, with a
bremsstrahlung spectrum with $kT=120\ud{30}{20} {\rm eV}$ and
\nhtt=0.6\ud{0.1}{0.2}.  A factor of 2 variability in X-ray flux over
a timescale of years is also found \citep{lira00}.  While previous
observers conclude this object is a ULX in NGC 247 based on its factor
of 2 variability and consistency with a high luminosity (\ee{39}
\cgslum) X-ray source at the distance of NGC 247, we cannot exclude
that it is a foreground INS due to the lack of an identified off-band
counterpart.  \unde.  \done\

{ \em 1RXS J020317.5-243832. } - \notanins\ (R03).  \done\
   
{ \em 1RXS J024528.9+262039. } -  \notanins\ (R03). \done\
  
{ \em 1RXS J025414.5+413530. } - This RASS/BSC source has a high
count rate (2.64\ppm0.05 PSPC c/s), and is extended on length scales of
$\sim$15\arcmin\.  In a 50 \ksec\ \chandra\ observation
(\obsid=908), an X-ray source extended on length scales of $\sim$5\arcmin\
is located $\sim$2.5\arcmin\ from the RASS/BSC position.   The next 
nearest RASS/BSC source is $\sim$42\arcmin\ away from the extended source.
The extended source has been identified as the galaxy cluster AWM~7
\citep{furusho03}.  We suggest therefore that this RASS/BSC source is 
AWM~7, and that the RASS/BSC astrometry for this object is incorrect 
by 2.6\arcmin.  \notanins. \done\

{ \em 1RXS J034414.1+240623. } - This X-ray source was redetected in
30\ksec\ of \rosat/HRI observations \citep{panzera03}, with a position
03\h44\m14\s.8,+24\deg06\arcmin05.7\arcsec, and an HRI count rate of
20\ppm1~\cpks, consistent with the detected RASS/BSC PSPC
count rate (64.5\ppm15.2~PSPC~\cpks; 24.3 \cpks\ expected for HRI).
The probability of chance association between the \rosat/HRI source
and a background AGN is \pchance=\ee{4.0}.   We therefore
conclude that the HRI source is associated with the RASS/BSC
source.

The closest catalogued USNO-B1 optical source is USNO-B1~1141-0042520 
($R=9.99$), located 3.8\arcsec\ away from the HRI source position. Within a
radius of 1000\arcsec\ we find 8 objects with as bright or brighter $R$
band magnitudes, producing a probability of chance spatial coincidence
of $\sim$\ee{-4}.  We therefore associate the RASS/BSC X-ray source
with this optical source, which is identified as BD+23~502. \notanins.

{ \em 1RXS J051541.7+010528. } - This X-ray source has been observed with 
\xmm\, and found to have an optical counterpart, identified as the 
long-period polar V1309 Orionis. \notanins\ \citep{schwarz05}. \done
  
{ \em 1RXS J053510.8-044850. } - Observed with \chandra/ACIS-I (\obsid=2549,
observation start time: 2002 August 26 13:49 UT, duration 48.8 \ksec).  

Based on the RASS/BSC count rate (200\ppm25 PSPC \cpks), the expected
\chandra/ACIS-I count rate is 164 \cpks\ which corresponds to a total
of $\sim$7940 counts detected during the \chandra\ observation from
this source.  

No individual point source with this count rate is found within 90\arcsec\ of
the RASS/BSC position (which is uncertain by 26\arcsec, 1 $\sigma$).
We calculate a 99\% confidence upper-limit to the flux
(i.e. count rate) of this X-ray source during the \chandra\
observation, assuming a point-source, of $<$0.29 \cpks, which
corresponds to a flux of $<$4.3\tee{-15} \cgsflux (0.1-2.4 keV), which
is approximately a factor of $2\times10^3$ fainter than the RASS/BSC
measured flux of 8.4\tee{-12}(0.1-2.4 keV).

There is a faint X-ray source located approximately 30\arcsec\
from the RASS/BSC position, with a count rate of 0.4\ppm0.1 \cpks\
(flux of $\sim$5.9\tee{-15} \cgsflux). The probability of chance 
association between the \chandra/ACIS-I source and a background AGN 
is \pchance=0.35.

 This RASS/BSC source is near an optical nebulosity
in the DSS.  A {\tt wavdetect} imaging analysis was performed with the
\chandra\ data, searching for sources on spatial scales of 2\arcsec,
4\arcsec, 8\arcsec, 16\arcsec, 32\arcsec, 64\arcsec, and 128\arcsec;
no spatially resolved X-ray source was found within 60\arcsec\ of the
RASS/BSC position at the flux level detected from the RASS/BSC source.
Thus, the nature of this RASS/BSC source cannot be determined. \unde. \done

{ \em 1RXS J053516.3-044033. } - Observed with \chandra/ACIS-I
(\obsid=2549, observation start time: 2002 August 26 13:49 UT, duration 48.8\ksec). 
A {\tt wavdetect} analysis finds a point source located at
05\h35\m41\s.2,-04\deg40\arcmin31\arcsec\ (\ppm0.06\arcsec
statistical, \ppm0.6\arcsec\ systematic).  This source was found to
have a count rate of 39.5\ppm0.9 \cpks, which corresponds to a flux of
5.8\tee{-13} \cgsflux (0.1-2.4 keV).  The RASS/BSC count rate
(52.4\ppm13.4 PSPC \cpks) corresponds to a flux to be 2.2\tee-12
\cgsflux (0.1-2.4 keV).  Thus, the \chandra\ source is found to be a
factor of $\times$4 fainter than the RASS/BSC source.
The probability of chance association between the \chandra/ACIS-I source
and a background AGN is \pchance=\ee{-4.8}.
 We therefore conclude that the \chandra\ X-ray source
is the RASS/BSC source, now fainter by a factor of $\sim$7.

2MASS~J053516.7-044032 is located 0.95\arcsec\ away from the 
\chandra\ source position.  The probability of chance association is 
\pchance=\ee{-2.9}. \notanins \done

{ \em 1RXS J053842.4-023525. } -
Observed with \chandra/HRC-I (\obsid=2560, duration 98.4\ksec).
A {\tt wavedetect} analysis detects two X-ray sources, 
3\arcsec\ apart,  with  positions 
05\h38\m44.\s7,$-$02\deg36\arcmin00\arcsec\ (\ppm0.001\arcsec\ statistical, 
\ppm0.6\arcsec\ systematic) and 05\h38\m44\s.8,-02\deg35\arcmin57\arcsec\
(\ppm0.01\arcsec\ statistical, \ppm0.6\arcsec\ systematic).
This binary is located 49 \arcsec\ from the RASS/BSC source position (which has an
uncertainty of $\ppm$11\arcsec).  Their position is 0.2\arcsec\ away from
the binary Sigma Orionis AB, USNO~J053844.7-023600.   The probability of chance 
association is \pchance=\ee{-2.9}, and we conclude the \chandra/HRI
source is the binary Sigma Orionis AB.  

The RASS/BSC count rate (638\ppm40 PSPC \cpks) corresponds to a flux
of 2.6\tee{-11}\cgsflux (0.1-2.4 keV).  We determine the \chandra
count rate, for both sources added together, to be 809\ppm2 \cpks,
which corresponds to a flux of 2.1\tee{-11} \cgsflux (0.1-2.4 keV).
The probability of chance association between the \chandra/HRC-I source
and a background AGN is \pchance=\ee{-4.7}.
We conclude the \chandra/HRC-I X-ray sources to be
1RXS~J053842.4-023525. \notanins.  \done

{ \em 1RXS J054042.8-020533. } - During 42\ksec\ of observations with 
\xmm/MOS-1 (\obsid=0101440301), an X-ray source is detected with a position
05\h40\m41\s.3,=02\deg05\arcmin38\arcsec\ (\ppm0.5\arcsec
statistical, \ppm2\arcsec\ systematic), 23\arcsec\ away from the 
RASS/BSC position.
It was found to have a count rate of 6.4 \ppm 0.5 \cpks, which
corresponds to a flux of 1.3\tee{-13}\cgsflux(0.1-2.4 keV).  The 
RASS/BSC source has a count rate of 177\ppm 22.1 \cpks, which 
corresponds to a flux of 7.5\tee{-12}\cgsflux(0.1-2.4 keV).
The probability of chance association between the \xmm\ source
and a background AGN is \pchance=\ee{-1.5}.

We note the presence of a source located at 
05\h40\m54\s.0,-02\deg03\arcmin02\arcsec, $\sim$3.8\arcmin\ from the
RASS/BSC position.  This source has a count rate of 72.0\ppm 1.3 \cpks,
which corresponds to a flux of 1.5\tee{-12}\cgsflux (0.1-2.4 keV).
The probability of chance association between the \xmm\ source
and a background AGN is \pchance=\ee{-1.8}.

Thus, we cannot conclude that either of  the two \xmm\ X-ray sources are 
associated with the RASS/BSC source. \unde.

{ \em 1RXS J054045.7-021119. } - A {\tt wavdetect} analysis was performed
 on a 7.4 \ksec\ duration \rosat/HRI observation (\obsid=rh201394n00)
An X-ray source was localized at 05\h40\m44\s.7,-02\deg11\arcmin54\arcsec,
($\ppm$1.4\arcsec, statistical, $\ppm$6\arcsec\ systematic), 37\arcsec\
from the RASS/BSC source position (which has a $\ppm$18\arcsec\ uncertainty).  

The RASS/BSC source count rate (161\ppm20 PSPC \cpks) corresponds to 
a flux of 6.8\tee{-12}\cgsflux(0.5-2.0 keV).
The HRI source was found to have a count rate of 4.47\ppm0.81 \cpks, which 
corresponds to a flux of 5.0\tee{-13}\cgsflux(0.1-2.4 keV).
The probability of chance association between the \rosat/HRI source
and a background AGN is \pchance=\ee{-2.8}.  We therefore
conclude that the HRI source is associated with the RASS/BSC
source.

A $B=13.48$  USNO-B1.0 source is located 4.6\arcsec\ away from the 
HRI source position, at 05\h40\m44.82\s,-02\deg11\arcmin50.0\arcsec.  For
objects as bright or brighter than $B=13.48$, the probability of
chance association is \ee{-3.0}. We conclude that the HRI object is
the RASS/BSC source, which has faded by a factor of $\sim$7.
\notanins.  \done

{ \em 1RXS J055228.1+155313. } - The \rosat/PSPC count rate is 
2.14\ppm0.07 counts \persec. No higher spatial resolution observations 
(\rosat/HRI, \chandra, or \xmm) exist of this source.  The nearest USNO-B1 object (USNOB1
1058-0090877) lies 6.9\arcsec\ away, and is a spectroscopically
identified white dwarf GD 71, which has been previously associated
with the RASS/BSC source \citep{fleming96}.  Based on the $R$-band
magnitude ($R=12.96$), and that there are 164 USNO-B1 objects as
bright or brighter within 1000\arcsec, the probability of chance
association is 7\tee{-3}.  \notanins. \done

{ \em 1RXS J055800.7+535358. } - This object has been associated with
an optical counterpart that is a known intermediate polar,  
with an X-ray period of 272.74 sec X-ray \citep{haberl94}.  \notanins. \done
 
{ \em 1RXS J060452.1-343331. } - During 21.3\ksec\ of observations with the
\rosat/HRI (\obsid=rh202301n00), an X-ray source is detected with a position   
06\h04\m52\s.11,-34\deg33\arcmin34.94\arcsec. (\ppm0.1\arcsec
statistical, \ppm6\arcsec\ systematic). 

The RASS/BSC source has a count rate of 428\ppm50 PSPC \cpks, which
corresponds to a flux of 8.8\tee{-12} \cgsflux (0.1-2.4 keV).  The
\rosat/HRI source was determined to have a count rate of 86.2 \ppm 2.0 \cpks,
which corresponds to a flux of 5.3\tee{-12} \cgsflux (0.1-2.4
keV). The probability of chance association between the \rosat/HRI source
and a background AGN is \pchance=\ee{-4.5}.

A $B=13.81$ USNO B1.0 source is located 1.2\arcsec\ away from the HRI source
position at 06\h04\m52\s.15,-34\deg33\arcmin36.0\arcsec.  The probability of chance association 
is \ee{-3.3}.  This optical star is
an eruptive variable V* AP Col, which has previously been associated
with the 1RXS~J060452.1$-$343331 \citep{fuhrmeister03}.
\notanins.  \done

{ \em 1RXS J063354.1+174612. } - This X-ray source is associated with
the radio-quiet pulsar SN 437 (Geminga).  It was associated based on
its X-ray pulsations, which have a period of 0.237 s.
\citep{halpern92}. For purposes of this paper, we treat this object
as an INS.   \done
 
{ \em 1RXS J074451.8+392733. } - This X-ray source has been observed with
\chandra\ and found to be significantly extended, and spatially
associated with a galaxy cluster \citep{ebeling07}. \notanins. \done
 
{ \em 1RXS J085247.0+223040. } - Performing a {\tt wavdetect} analysis on \xmm/MOS-1
\obsid=0144500101 (0.7 \ksec\ duration), an X-ray source was localized at
08\h52\m44\s.6,+22\deg30\arcmin52\arcsec\ (\ppm1.00\arcsec
statistical, \ppm2\arcsec\ systematic).

The \xmm\ source was found to have a count rate of 71\ppm17 \cpks,
which corresponds to a flux of 8.6\tee{-13}\cgsflux (0.1-2.4 keV).  The
RASS/BSC source count rate is 165\ppm25 PSPC \cpks, or
3.2\tee{-12}\cgsflux (0.1-2.4 keV).  
The probability of chance association between the \xmm source
and a background AGN is \pchance=\ee{-2.8}

A USNO-B1.0 source, USNO~J085244.7+223054, is located 2.2\arcsec\ 
away from the \xmm\ position.
For objects as bright or brighter than $B=14.61$, the probability of
chance association is \ee{-3.9}.    \notanins.  \done

{ \em 1RXS J091112.2+174634. } - This X-ray source has been
observed with \chandra\  and found to be significantly
extended, and spatially associated with a cluster of galaxies
\citep{ebeling07}. \notanins.  \done

{ \em 1RXS J101628.3-052026. } - Previously identified with a
spectroscopically classified white dwarf \citep{fleming96}, at
10\h16\m28\s.6-05\deg20\arcmin27\arcsec, with $B=13.3$, $V=14.3$, $R=12.02$
(USNO-B1.0 0846-0208321) and an approximately co-local 2MASS source
2MASS~J101628.67$-$052032.0 with $(J=10.607, H=9.99, K=9.77)$.  The
2MASS source is 7\arcsec\ from the RASS/BSC position, and there are 7 such
objects as bright or brighter as $J=10.607$ within 1000\arcsec, making the
probability of chance positional coincidence $\sim$3\tee{-4}.
\notanins. \done

{ \em 1RXS J102954.3+614732. } - This object was observed with
\rosat/HRI (obsid rh704085n00, 4.9 \ksec).  Two point sources lie
1\arcmin\ and 1.5\arcmin\ from the RASS/BSC position, with no other
detected X-ray sources within 3\arcmin.  The position of the closer
HRI X-ray source is 10\h29\m50\s.208,+61\deg46\arcmin45.65\arcsec
(\ppm 1\arcsec\ statistical, \ppm 6\arcsec\ systematic). 

The RASS/BSC source has a count rate of 145\ppm18 PSPC \cpks, which
corresponds to a flux of 1.4\tee{-12} \cgsflux (0.1-2.4 keV),
assuming \nh=7.8\tee{19} \perval{cm}{-2} and photon power-law slope of
$\alpha=2$.  The HRI source was determined to have a count rate of 18.3
\ppm 1.8 \cpks, which corresponds to a flux of 7.6\tee{-13}.
\cgsflux (0.1-2.4 keV).  
The probability of chance association between the \rosat/HRI source
and a background AGN is \pchance=\ee{-2.6}.

The factor of $\sim\times2$ 
difference in fluxes is considered to  to be consistent with 
spectral uncertainties and
source variability, and we therefore associate the HRI X-ray source
with the RASS/BSC X-ray source.   

The optical
source USNO~J102950.97+614643.2 ($B=17.35$) is  located 5.9\arcsec\ away from the HRI
source position.  The probability of chance association with an
object as bright or brighter than $B=17.35$ is \ee{-2.3}.  
\notanins.  \done 

{ \em 1RXS J103347.4-114146. } - There have been no X-ray observations
with the specified instrumentation.  \unde.  \done
 
{ \em 1RXS J104710.3+633522. } - Source has been previously found
(R03) to be a magnetic cataclysmic variable 
\citep{singh95a}.  \notanins. \done

{ \em 1RXS J115309.7+545636.} - This object was found to exhibit 
variability in the X-ray band, being undetected at a flux limit 
lower than that of the RASS/BSC detection (R03). \unde.  \done
 
{ \em 1RXS J123319.0+090110. } - This X-ray source has been found  to be a
DM star  \citep[R03;][]{marino00}. \notanins  \done  

{ \em 1RXS J130547.2+641252. } - Observed with HRI but undetected;
therefore, excluded previously as an INS on the basis of X-ray
variability (R03).  However, in the present analysis,
X-ray variability is not grounds for exclusion as an INS candidate.

\xmm/PN \obsid=0151790701 (6.9 \ksec\ duration) was examined.
 Performing a {\tt wavedetect} analysis revealed that the
X-ray point source closest to the RASS/BSC position is 175\arcsec
away, at 13\h06\m01\s.6,+64\deg10\arcmin27\arcsec. The RASS/BSC X-ray
source positional uncertainty is $\ppm$9\arcsec. 

The RASS/BSC count rate (167\ppm20 PSPC \cpks) corresponds to a flux of 
2.3\tee{-12} \cgsflux (0.1-2.4 keV).
The \xmm\ source was determined to have a count rate of 15.6 \ppm 1.6
\cpks, which corresponds to a flux of 4.2\tee{-14} \cgsflux (0.1-2.4
keV).
The probability of chance association between the \xmm\ source
and a background AGN is \pchance=0.45.  We cannot 
conclude that the two sources are related.

We calculate a 99\% confidence upper-limit to the flux of
this X-ray source during the \xmm\ observation, assuming a point source
of $<$2.0 \cpks, which corresponds to a flux of $<$5.3 \tee{-15}
\cgsflux (0.1-2.4 keV), which is a approximately a factor of
400$\times$ fainter than the RASS/BSC measured flux of
2.3\tee{-12}(0.1-2.4 keV).  \unde. \done
  
{ \em 1RXS J130753.6+535137. } - \notanins\ (R03).  \done\ 

{ \em 1RXS J130848.6+212708. } - Previously identified INS
\citep{hambaryan02}.  \done
 
{ \em 1RXS J132833.1-365425. } - \notanins. (R03) \done
  
{ \em 1RXS J134210.2+282250. } - Identified as a CV in M3
\citep[R03;][]{dotani99}.  \notanins.  \done.

{ \em 1RXS J145010.6+655944. } - \notanins\ (R03). \done
 
{ \em 1RXS J145234.9+323536. } - This object was previously found to
found to exhibit variability in the X-ray band(R03), 
which in the present analysis is no longer grounds
for exclusion as an INS candidate.  Observed with \chandra, but no 
X-ray source was detected (R03).  \unde.
\done
  
{ \em 1RXS J160518.8+324907. } - Previously identified INS \citep{motch99}. INS.  \done
 
{ \em 1RXS J161455.4-252800. } -Performing a {\tt wavdetect} analysis 
on \rosat/HRI observation (\obsid=rh201993n00, duration 3.25\ksec), we 
find a source 2.4\arcsec\ away from the RASS/BSC position, at 
16\h14\m55\s.3,-25\deg27\arcmin57\arcsec\ (\ppm0.8\arcsec\ statistical, \ppm6\arcsec\ 
systematic).  This HRI source was found to have a count rate 
of 37\ppm4 \cpks, which corresponds to a flux of 
3.8\tee{-12}\cgsflux (0.1-2.4 keV). The RASS/BSC count rate is 
100\ppm20 \cpks, which corresponds to a flux of 
3.9\tee{-12}\cgsflux (0.1-2.4 keV).  
The probability of chance association between the \rosat/HRI source
and a background AGN is \pchance=\ee{-6.3}.
We conclude
that the two sources are associated. 

USNO~J161455.4-252800 lies 
3.2 arcseconds away from the HRI position.  For objects as 
bright or brighter than $R=9.33$, the probability of chance 
association is \ee{-4.2}. \notanins.  \done

{ \em 1RXS J162721.6-244144. } - Observations using \chandra/ACIS-I of the
rho Ophiuchi region \citep{imanishi01} localize the nearest X-ray
source to 16\h27\m21\s.4,-24\deg41\arcmin43\arcsec,
(\ppm0.9\arcsec, 90\% confidence).   The RASS/BSC count rate
(93.7\ppm17.8 PSPC \cpks) corresponds to an expected \chandra\ ACIS-I
count rate of 137 \cpks.   \chandra\ observation photon count
rate for this source is 1.39 \cpks, a factor of 100$\times$ fainter
than the expected count rate.

There is a much brighter \chandra\ X-ray source 29\arcsec
away from the RASS/BSC position (which has $\ppm$14\arcsec\ uncertainty), 
localized at 16\h27\m19\s.5,-24\deg41\arcmin40\arcsec.  
This source was found to have a count rate of
83.8 \cpks \citep{imanishi01}, a factor of 0.6$\times$ the expected
ACIS-I count rate. 
The probability of chance association between the \chandra/ACIS-I source
and a background AGN is \pchance=\ee{-3.5}.
We conclude that the two sources are associated. 

There is a nearby USNO source, USNO~J162719.5-244140, 0.4\arcsec\
away from the position of the brighter \chandra\ source.  The
probability of chance association is \pchance=\ee{-3.3}. Thus,
we conclude that 1RXS~J162721.6-244144=CXO
J162719.5$-$244140.6=USNO~J162719.5-244140.  \notanins.  \done
 
{ \em 1RXS J163212.8-244013. } - A {\tt wavdetect} analysis 
was performed with \rosat/HRI observation (\obsid=rh201839n00, duration 6.33 \ksec).
An X-ray source was localized at 16\h32\m11\s.9,-24\deg40\arcmin21\arcsec\
($\ppm$1.3\arcsec\ statistical, $\ppm$6\arcsec\ systematic), 14\arcsec
from the RASS/BSC source position (which has uncertainty of $\ppm$10\arcsec).

Based on the RASS/BSC count rate (78.6\ppm15.8 PSPC \cpks), the 
RASS/BSC source has a flux of 2.8\tee{-12} \cgsflux (0.1-2.4 keV)

The HRI source was found to have a count rate of 24.6\ppm2.0 \cpks, which
corresponds to a flux of 2.3\tee{-12} \cgsflux (0.1-2.4 keV).  
The probability of chance association between the \rosat/HRI source
and a background AGN is \pchance=\ee{-4.5}.

A $B=15.27$ USNO-B1.0 source, USNO~J163211.8-244021, is located  
1.7 arcseconds from HRI position.  For objects as bright or brighter 
than $B=15.27$, the probability of chance association is \ee{-3.6}.  \notanins.  \done

{ \em 1RXS J163421.2+570933. } -  \notanins. (R03) \done
 
{ \em 1RXS J163910.7+565637. } - \notanins. (R03)  \done

{ \em 1RXS J173157.7-335007. } - This source has been identified with
a known and well-studied LMXB, MXB 1728-34 \citep{liu01}.
\notanins. \done

{ \em 1RXS J173253.6-371200. } -This object has no X-ray observations
with the specified instrumentation. \unde.  \done

{ \em 1RXS J173319.3-255416. } - There is a RASS/BSC source 15\arcmin
away (by visual inspection), 1RXS J173413.0-260527, also known as
KS~1731-261, a catalogued LMXB.  By visual inspection of the RASS image,
there does not appear to be any independent X-ray source at the given
position for 1RXS~J173319.3$-$255416, which is in the PSF wing of
KS~1731-261.  

Over 200\ksec\ of integration with \xmm\ have been accumulated in which
the RASS/BSC position is 16\arcmin\ off-axis, and 90\arcsec\ outside
the field-of-view (the observations target KS~1731-261).  

We then examined two \rosat/HRI observations of the LMXB, \obsid=rh400718n00 (1.08 \ksec)
and \obsid=rh400718a01 (4.68 \ksec).  The RASS/BSC count rate  (59.2 \ppm 17.2 \cpks) 
corresponds to a flux of 8.7\tee{-13} \cgsflux (0.1-2.4 keV). Thus, the 
expected \rosat/HRI count rate is 222 \cpks.  

For both observations, we find no such bright individual point sources within 
90\arcsec\ of the RASS/BSC position (which is uncertain by 14\arcsec\, 1$\sigma$).

For \obsid=rh400718n00, we calculate a 99\% confidence upper-limit to the flux of 
this X-ray source during the \rosat/HRI observation, assuming a point source of
$<$10.2 \cpks, which corresponds to a flux of $<$3.98\tee{-13} \cgsflux 
(0.1-2.4 keV), which is approximately a factor of $2\times$ fainter
than the RASS/BSC measured flux.

Our conclusion is that this RASS/BSC catalog entry is due to a
spurious source detection in the wing of KS~1731-260. \notanins. \done

{ \em 1RXS J173546.9-302859. } - This X-ray source has been associated
with a neutron star X-ray transient X1732-304 in quiescence in the
globular cluster Terzan~1 \citep{wijnands02}. \notanins. \done
 
{ \em 1RXS J180132.3-203132. } - This X-ray source has been associated
with a well-studied LMXB, X~SGR~X-3 (4U ~1758$-$20), including recent
simultaneous X-ray optical observations \citep{kong06}.  \notanins.
\done
 
{ \em 1RXS J181506.1-120545. } - This X-ray source has been associated
with a faint LMXB (XB~1812-12).  A \chandra\ observation reveals an
object which is 2.4\arcsec\ offset from the RASS/BSC position,
(18\h15\m06\s.1,-12\deg05\arcmin47\arcsec) with a flux of
4.4\tee{-10}\cgsflux(1-10keV) \citep{wilson03}, which is comparable to
the flux of the RASS/BSC source (1.3\ppm0.1 PSPC c \persec, corresponding to
9.3\tee{-11} \cgsflux)   
The probability of chance association between the \chandra\ source
and a background AGN is \pchance=\ee{-9.2}.
We therefore conclude that the two objects are associated.  \notanins. \done

{ \em 1RXS J182102.0-161309. } - No optical counterpart was found in
USNO-A2, although the DSS image was found to contain a nebulosity
\citep{rutledge00}.  

This X-ray source has been observed with \xmm\ (\obsid=0144500101,
observation date 2003 March 11, duration 35\ksec, 8.9\ksec\ after flare
removal), approximately 7.4\arcmin\ off-axis.

The RASS/BSC count rate (161\ppm28 PSPC \cpks) corresponds to a flux
of 2.9\tee{-11} \cgsflux (0.1-2.4 keV).  Thus, the expected
\xmm\ count rate is 1440 \cpks, which corresponds to a total of
$\sim$12816 counts detected during the \xmm\ observation of this
source.

A {\tt wavdetect} analysis, searching for sources on spatial scales of
4\arcsec, 8 \arcsec, 16\arcsec, 32\arcsec, 64\arcsec, and 128 \arcsec
of the PN image localized one X-ray sources located 150\arcsec\ from the
RASS/BSC position (which has an uncertainty of $\ppm$37\arcsec)

The brightest nearby source was found to have a count rate of
3.9\ppm1.1 \cpks\ which corresponds to a flux of
3.3\tee{-14}\cgsflux(0.5-2 keV). 
The probability of chance association between the \xmm\ source
and a background AGN is \pchance=0.10.  This does not
support a  conclusion that the brightest \xmm\ source is 
associated with the RASS/BSC source.

We calculate a 99\% confidence upper-limit to the flux of this X-ray 
source during the \xmm\ observation, assuming a point source of 
$<$1.69 \cpks, which corresponds to a flux of $<$3.4 \tee{-14} \cgsflux 
(0.1-2.4 keV), which is approximately a factor of  $\times$800 
 fainter than the RASS/BSC measured flux.  \unde. \done

{ \em 1RXS J183543.6-325928. } - This X-ray source has been associated
with an LMXB (XB~1832-330) in the globular cluster NGC
6652. \citep{liu01} \notanins.  \done
 
{ \em 1RXS J185557.3+233410. } - This object has no X-ray observations
with the specified instrumentation. \unde.  \done
 
{ \em 1RXS J191426.1+245641. } - This object has been associated with
a cataclysmic variable V*~V407~Vul \citep{ramsay00},
with extensive \chandra\ and \xmm\ observational history.  \notanins.
\done

{ \em 1RXS J221144.6-034947. } - 
In an 80\ksec\ \chandra\ observation (\obsid=3284), an X-ray 
source is localized using {\tt wavdetect}, at 22\h11\m45\s.8,-03\deg49\arcmin48\arcsec, 
with $\ppm$0.15\arcsec\ statistical, $\ppm$0.6\arcsec
systematic uncertainty, located $\sim$ 18\arcsec\ from the RASS/BSC position.   
The {\tt wavdetect} analysis indicates that this source has a PSF ratio
of 5.81, so we consider it to be extended. 
The RASS/BSC source has been identified as galaxy cluster 
RXCJ2211.7-0350 \citep{bohringer04}.

\notanins.  \done

{ \em 1RXS J221403.0+124207. } - This X-ray source has been associated
with an historical dwarf nova RU Peg \citep{silber94}.  While no X-ray
observations with the specified instrumentation have been analysed,
one such observation with \xmm\ was taken in 2008 June 09 can be used to
confirm the localization; we tentatively conclude this X-ray source is
the dwarf nova RU Peg.  \notanins \done

{ \em 1RXS J223832.0-151809. } - A {\tt wavedetect} analysis on a
 pointed \rosat/PSPC observation (\obsid=RP201723N00; 13.2\ksec\
 duration) determines the X-ray position to be
 22\h38\m32\s.040,-15\deg18\arcmin03\arcsec.91 (\ppm 0.4\arcsec\
 statistical, 6\arcsec\ systematic).

Archival images in DSS and 2MASS (1982, 1987, 1991, and 2000) reveal a
high proper-motion star which was located within the X-ray error
circle at the time of the RASS/PSPC observation.  
2MASS~J223833.7-151757 ($J=6.6$) is located 25\arcsec\ away from the PSPC
source position.  The probability of chance association for this object is
\ee{-3.7}.  \notanins.

{ \em 1RXS J230334.0+152019. } - We analysed a \rosat/HRI
observation (\obsid=rh701842n00, 7.3 \ksec\ duration).  A {\tt wavdetect}
imaging analysis was performed, searching for sources on spatial
scales of 2\arcsec, 4\arcsec, 8\arcsec, 16\arcsec, 32\arcsec, and
64\arcsec;
a source was found 24\arcsec\ away from the RASS/BSC position, at
23\h03\m32\s.3,+15\deg20\arcmin16\arcsec\ ($\ppm$0.5\arcsec\ statistical,
$\ppm$6\arcsec\ systematic).

The RASS/BSC count rate (60.2\ppm13.7 PSPC \cpks) corresponds to a flux of  
1.5\tee{-12}\cgsflux (0.1-2.4 keV).  We determine the \rosat/HRI count rate
to be 9.6\ppm1.2 \cpks, corresponding to a flux of 6.9\tee{-13}\cgsflux
(0.1-2.4 keV), a factor of $\times$2 fainter than the RASS/BSC flux.  
The probability of chance association between the \rosat/HRI source
and a background AGN is \pchance=\ee{-3.3}.
We therefore conclude that the HRI source
is the RASS/BSC source, now fainter by a factor of $\sim$2.

The closest catalogued USNO-B1 optical source is USNO~J230332.4+152009 ($B=20.4$), 
is located 6.3\arcsec\ away from the HRI position.  For objects as bright or brighter than $B=20.4$, 
the probability of chance association is \ee{-1.4}.  \unde  \done\

{ \em 1RXS J233407.2-033533. } - There have been no X-ray observations
with the specified instrumentation.  \unde.  \done
 
{ \em 1RXS J233757.2+271031. } - We examined the \xmm/MOS-1
observation (\obsid=0002960101, 11.5 \ksec).  A {\tt wavdetect}
imaging analysis was performed, searching for sources on spatial
scales of 2\arcsec, 4\arcsec, 8\arcsec, 16\arcsec, 32\arcsec,
64\arcsec, and 128\arcsec; a source was found 57\arcsec\ away from the
RASS/BSC position, at 23\h37\m54\s.8,+27\deg11\arcmin10\arcsec
($\ppm$0.9\arcsec\ statistical, $\ppm$2\arcsec\ systematic).

The RASS/BSC count rate (68.4\ppm12.9 PSPC \cpks) corresponds to a
flux of 1.7\ee{-12}\cgsflux (0.1-2.4 keV).  We find the \xmm/MOS-1
source count rate to be 34.6 \ppm 2.1 \cpks, which corresponds to a
flux of 4.5\tee{-13}\cgsflux (0.1-2.4 keV), a factor of $\times$4
fainter than the RASS/BSC flux.
The probability of chance association between the \xmm\ source
and a background AGN is \pchance=\ee{-2.3}.
We therefore conclude that the \xmm\ X-ray
source is the RASS/BSC source, now fainter by a factor of $\sim$4.

The \xmm\ source is located 9.5\arcsec\ away from USNO~J233756.0+271117
($B=18.46$).  The probability of chance association is
\pchance=\ee{-1.4}.  The \xmm\ source is also located 11.1\arcsec\ away
from 2MASS~J233755.5+271113 ($J=15.7$).  The probability of chance
association is $\pchance=\ee{-1.3}$.
 \unde.  \done



\newpage

\begin{deluxetable}{lcllcrll}
\rotate
\tablewidth{0pt}
\tabletypesize{\small}
\tablecolumns{8}
\label{tab:sources}

\tablecaption{Candidates and Identifications}

\tablehead{
  \colhead{1RXSJ}   &
  \colhead{\pnoid}  &
  \colhead{Type}    &
  \colhead{Name}    &
  \colhead{Observed?\tnm{a}} &
  \colhead{Hardness Ratio}    &
  \colhead{ID}      }

\startdata
J001832.0$+$162634 &  0.872 &              Cluster of Galaxies &               ClG 0015.9+1609 &  \pobs{CXH} &    0.73 &   \notanins &         \\ 
J003854.9$-$034252 &  0.818 &                          \nodata &                       \nodata &    \pobs{S} &    0.56 &   \notanins &         \\ 
J004330.9$+$411452 &  0.865 &                          \nodata &                2E 0040.8+4057 & \pobs{CXHS} &    0.63 &   \notanins &         \\ 
J004704.8$-$204743 &  0.855 &                          \nodata &                       \nodata &   \pobs{XH} &    1.00 &       \unde &         \\ 
J012428.1$-$335504 &  0.861 &          High proper-motion Star &                     G 269-153 &    \pobs{S} & $-$0.14 &   \notanins &         \\ 
J013653.5$-$351012 &  0.884 &                 Seyfert 1 Galaxy &        2MASSI J0136544-350952 &  \pobs{XHS} & $-$0.81 &   \notanins &         \\ 
J014205.0$+$213045 &  0.858 &              Cluster of Galaxies &                ClG J0142+2131 &    \pobs{S} &    1.00 &   \notanins &         \\ 
J015311.6$-$210545 &  0.876 &          High proper-motion Star &                     G 274-113 &    \pobs{S} & $-$0.14 &   \notanins &         \\ 
J020146.5$+$011717 &  1.000 &                             Star &               RX J0201.7+0117 &    \pobs{S} & $-$0.25 &   \notanins &         \\ 
J020210.5$-$020616 &  0.911 &                          \nodata &                       \nodata &    \pobs{S} &    0.04 &       \unde &         \\ 
J020317.5$-$243832 &  0.939 &                          \nodata &                       \nodata &    \pobs{C} & $-$0.16 &   \notanins &         \\ 
J020503.6$-$173717 &  0.815 &                          \nodata &                       \nodata &    \pobs{S} & $-$0.52 &       \unde &         \\ 
J022152.6$+$281047 &  0.833 &                          \nodata &               RX J0221.8+2810 &    \pobs{S} & $-$0.15 &   \notanins &         \\ 
J024528.9$+$262039 &  0.929 &                             Star &                       \nodata &    \pobs{C} & $-$0.19 &   \notanins &         \\ 
J024814.8$-$193954 &  0.808 &                          \nodata &                       \nodata &    \pobs{S} & $-$0.19 &   \notanins &         \\ 
J024946.0$-$382540 &  1.000 &                          \nodata &                       \nodata &    \pobs{S} & $-$0.24 &   \notanins &         \\ 
J025414.5$+$413530 &  0.801 &                          \nodata &                       \nodata &  \pobs{CXH} &    0.99 &   \notanins &         \\ 
J031413.7$-$223533 &  0.925 &                   Nova-like Star &                     V* EF Eri &   \pobs{XS} & $-$0.61 &   \notanins &         \\ 
J032620.8$+$113106 &  1.000 &                  T Tau-type Star & [LHC2000] J032621.24+113055.9 &    \pobs{S} & $-$0.16 &   \notanins &         \\ 
J033642.5$-$095506 &  0.816 &                          \nodata &                       \nodata &    \pobs{S} &    0.71 &       \unde &         \\ 
J034414.1$+$240623 &  0.939 &                  Star in Cluster &                   BD+23   502 &    \pobs{H} &    0.79 &   \notanins &         \\ 
J035101.5$+$141404 &  0.833 &                          \nodata &                       \nodata &    \pobs{S} & $-$0.35 &   \notanins &         \\ 
J040314.6$-$360927 &  0.832 &                             Star &              2E 0401.4$-$3617 &    \pobs{S} & $-$0.22 &   \notanins &         \\ 
J040543.6$-$282114 &  0.807 &                          \nodata &                       \nodata &    \pobs{S} & $-$0.24 &       \unde &         \\ 
J040814.3$+$400723 &  0.811 &                          \nodata &                       \nodata &    \pobs{S} & $-$0.14 &   \notanins &         \\ 
J040817.9$+$294951 &  0.825 &                          \nodata &                       \nodata &    \pobs{S} &    1.00 &   \notanins &         \\ 
J040913.8$+$110833 &  0.905 &                          \nodata &                       \nodata &    \pobs{S} &    1.00 &   \notanins &         \\ 
J041215.8$+$644407 &  1.000 &                       Flare Star &                      GJ  3266 &    \pobs{S} & $-$0.51 &   \notanins &         \\ 
J042513.4$+$171548 &  0.850 &                      White Dwarf &                   V* V805 Tau &    \pobs{S} & $-$0.30 &   \notanins &         \\ 
J043334.8$+$204437 &  0.925 &                       Flare Star &                      GJ  3296 &    \pobs{S} & $-$0.15 &   \notanins &         \\ 
J044048.0$+$292440 &  0.914 &                          \nodata &                       \nodata &    \pobs{S} &    0.93 &   \unde &         \\ 
J050909.9$+$152740 &  0.890 &                       Flare Star &                      GJ  3335 &    \pobs{S} & $-$0.13 &   \notanins &         \\ 
J051028.9$+$022051 &  0.805 &                          \nodata &                       \nodata &    \pobs{S} &    0.67 &   \notanins &         \\ 
J051315.9$+$025227 &  0.925 &                          \nodata &                       \nodata &    \pobs{S} &    0.37 &   \notanins &         \\ 
J051319.1$+$013525 &  0.855 &                          \nodata &                       \nodata &    \pobs{S} & $-$0.14 &   \notanins &         \\ 
J051354.0$+$023722 &  0.816 &                          \nodata &                       \nodata &    \pobs{S} &    0.00 &   \notanins &         \\ 
J051541.7$+$010528 &  0.961 &     Cataclysmic Var. AM Her type &                  V* V1309 Ori &   \pobs{XH} & $-$0.92 &   \notanins &         \\ 
J051723.3$-$352152 &  0.912 &          High proper-motion Star &                      L  449-1 &    \pobs{S} & $-$0.18 &   \notanins &         \\ 
J051854.5$+$323827 &  1.000 &                          \nodata &                       \nodata &    \pobs{S} &    0.06 &   \notanins &         \\ 
J052929.6$+$031838 &  0.805 &                          \nodata &                       \nodata &    \pobs{S} &    0.69 &   \notanins &         \\ 
J053510.8$-$044850 &  1.000 &         Nebula of unknown nature &                       \nodata &  \pobs{CXH} &    1.00 &       \unde &         \\ 
J053516.3$-$044033 &  0.903 &                          \nodata &              2E 0532.8$-$0442 &  \pobs{CXH} &    1.00 &   \notanins &         \\ 
J053842.4$-$023525 &  0.815 &                          \nodata &                       \nodata &  \pobs{CXH} &    0.51 &   \notanins &         \\ 
J054042.8$-$020533 &  0.808 &                          \nodata &                       \nodata &  \pobs{CXH} &    0.67 &       \unde &         \\ 
J054045.7$-$021119 &  1.000 &                    Variable Star &                   V* V611 Ori &   \pobs{XH} &    0.82 &   \notanins &         \\ 
J055228.1$+$155313 &  0.815 &                      White Dwarf &                       GD   71 &     \nodata & $-$0.99 &   \notanins &         \\ 
J055734.5$-$273534 &  0.889 &                          \nodata &                       \nodata &    \pobs{S} &    0.46 &       \unde &         \\ 
J055800.7$+$535358 &  0.849 &     Cataclysmic Var. DQ Her type &                   V* V405 Aur &   \pobs{XH} & $-$0.53 &   \notanins &         \\ 
J060452.1$-$343331 &  0.961 &           Eruptive variable Star &                     V* AP Col &    \pobs{H} &    0.06 &   \notanins &         \\ 
J062203.6$+$744106 &  0.826 &                          \nodata &                       \nodata &    \pobs{S} &    0.32 &   \notanins &         \\ 
J063354.1$+$174612 &  0.871 &                           Pulsar &                       SN  437 &  \pobs{CXH} & $-$0.92 &         INS &         \\ 
J074451.8$+$392733 &  0.813 &                          \nodata &                       \nodata &    \pobs{C} &    0.83 &   \notanins &         \\ 
J075523.9$+$372634 &  0.907 &                  Possible BL Lac &      SDSS J075523.11+372618.8 &    \pobs{S} &    0.96 &   \notanins &         \\ 
J075556.7$+$832310 &  0.904 &                       Flare Star &                      GJ  1101 &    \pobs{S} & $-$0.21 &   \notanins &         \\ 
J082124.5$-$362848 &  0.847 &                          \nodata &                       \nodata &    \pobs{S} &    0.52 &   \notanins &         \\ 
J082355.1$+$394745 &  0.832 &                          \nodata &               RX J0823.9+3947 &    \pobs{S} &    0.77 &   \notanins &         \\ 
J084127.7$-$102843 &  0.836 &                          \nodata &                       \nodata &    \pobs{S} &    1.00 &       \unde &         \\ 
J085247.0$+$223040 &  0.961 &                          \nodata &               RX J0852.7+2230 &    \pobs{X} &    0.13 &   \notanins &         \\ 
J090137.2$-$052341 &  0.815 &                          \nodata &                       \nodata &    \pobs{S} & $-$0.11 &   \notanins &         \\ 
J090717.4$+$225254 &  0.849 &                             Star &                     HD  78141 &    \pobs{S} & $-$0.06 &   \notanins &         \\ 
J091010.2$+$481317 &  0.931 &                 Seyfert 1 Galaxy &                 QSO B0906+484 &    \pobs{S} & $-$0.43 &   \notanins &         \\ 
J091112.2$+$174634 &  0.849 &                          \nodata &               RX J0911.2+1746 &    \pobs{C} &    0.46 &   \notanins &         \\ 
J094432.8$+$573544 &  1.000 &             BL Lac - type object &      SDSS J094432.32+573536.0 &   \pobs{HS} & $-$0.39 &   \notanins &         \\ 
J094454.2$-$122047 &  0.804 &          High proper-motion Star &                      G 161-71 &    \pobs{S} & $-$0.15 &   \notanins &         \\ 
J094831.1$-$333814 &  0.801 &                          \nodata &                       \nodata &    \pobs{S} &    0.73 &       \unde &         \\ 
J101628.3$-$052026 &  0.835 &                      White Dwarf &               GSC 04910-01132 &     \nodata & $-$1.00 &   \notanins &         \\ 
J102213.7$+$735433 &  0.803 &                          \nodata &               RX J1022.2+7354 &    \pobs{S} &    0.09 &   \notanins &         \\ 
J102954.3$+$614732 &  0.896 &                          \nodata &               RX J1029.9+6147 &    \pobs{H} & $-$0.29 &   \notanins &         \\ 
J103347.4$-$114146 &  0.814 &                          \nodata &                       \nodata &     \nodata & $-$0.96 &       \unde &         \\ 
J104710.3$+$633522 &  0.939 &     Cataclysmic Var. DQ Her type &                     V* FH UMa &    \pobs{X} & $-$0.97 &   \notanins &         \\ 
J110521.6$-$073525 &  0.870 &                          \nodata &             RX J1105.3$-$0735 &    \pobs{S} & $-$0.13 &   \notanins &         \\ 
J112430.5$+$435131 &  0.832 &                          \nodata &               RX J1124.5+4351 &    \pobs{S} &    0.84 &   \notanins &         \\ 
J115309.7$+$545636 &  1.000 &                          \nodata &               RX J1153.1+5456 &    \pobs{C} & $-$1.00 &       \unde &         \\ 
J120711.0$+$364745 &  0.876 &                 Seyfert 1 Galaxy &       2MASS 12071782+3648008 &    \pobs{S} & $-$0.03 &   \notanins &         \\ 
J121732.6$+$152843 &  0.961 &                          \nodata &               RX J1217.5+1528 &    \pobs{S} &    0.45 &   \notanins &         \\ 
J121900.7$+$110727 &  0.808 &                          \nodata &               RX J1219.0+1107 &    \pobs{S} & $-$0.25 &   \notanins &         \\ 
J122308.4$+$110054 &  0.953 &                          \nodata &               RX J1223.1+1100 &    \pobs{S} &    0.48 &   \notanins &         \\ 
J122940.6$+$181645 &  1.000 &             BL Lac - type object &  [ZEH2003] RX J1229.6+1816  1 &   \pobs{CS} & $-$0.38 &       \unde &         \\ 
J123319.0$+$090110 &  0.933 &            Star in double system &                    GJ   473 B &     \nodata & $-$0.43 &   \notanins &         \\ 
J124849.0$+$333454 &  0.845 &                          \nodata &               RX J1248.8+3334 &    \pobs{S} & $-$0.48 &   \notanins &         \\ 
J125015.2$+$192357 &  1.000 &                 Seyfert 1 Galaxy &  [ZEH2003] RX J1250.2+1923  1 &    \pobs{S} & $-$0.64 &   \notanins &         \\ 
J125721.8$+$365431 &  0.810 &                          \nodata &               RX J1257.3+3654 &    \pobs{S} &    0.79 &       \unde &         \\ 
J125947.9$+$275636 &  0.961 &              Cluster of Galaxies &                     ACO  1656 & \pobs{CXHS} &    0.33 &   \notanins &         \\ 
J130034.2$+$054111 &  0.954 &                       Flare Star &                     V* FN Vir &    \pobs{S} & $-$0.16 &   \notanins &         \\ 
J130205.2$+$155122 &  1.000 &                          \nodata &               RX J1302.0+1551 &    \pobs{S} &    0.13 &       \unde &         \\ 
J130402.8$+$353316 &  1.000 &                           Quasar &                QSO B1301+358A &    \pobs{S} & $-$0.34 &   \notanins &         \\ 
J130547.2$+$641252 &  1.000 &                             Star &               RX J1305.7+6412 &   \pobs{XH} & $-$0.82 &       \unde &         \\ 
J130631.3$+$192229 &  1.000 &                          \nodata &               RX J1306.5+1922 &    \pobs{S} &    0.11 &   \notanins &         \\ 
J130753.6$+$535137 &  0.939 &     Cataclysmic Var. AM Her type &                     V* EV UMa &   \pobs{XH} & $-$0.90 &   \notanins &         \\ 
J130848.6$+$212708 &  1.000 &           Neutron Star Candidate &               RX J1308.8+2127 &  \pobs{CXH} & $-$0.20 &         INS &         \\ 
J131011.9$+$474521 &  0.960 &          High proper-motion Star &                     LHS  2686 &    \pobs{S} &    0.25 &   \notanins &         \\ 
J132041.2$-$030010 &  0.815 &                          \nodata &                       \nodata &    \pobs{S} & $-$1.00 &       \unde &         \\ 
J132833.1$-$365425 &  0.961 &          High proper-motion Star &                       \nodata &    \pobs{C} & $-$0.57 &   \notanins &         \\ 
J133032.3$+$720931 &  0.961 &                          \nodata &               RX J1330.5+7209 &    \pobs{S} &    0.18 &   \notanins &         \\ 
J133825.0$-$251634 &  0.872 &                          \nodata &                 [FS2003] 0678 &    \pobs{S} &    0.10 &   \notanins &         \\ 
J134210.2$+$282250 &  1.000 &        Cataclysmic Variable Star &               RX J1342.1+2822 &   \pobs{CH} & $-$0.95 &   \notanins &         \\ 
J134619.1$-$141834 &  0.960 &                          \nodata &                       \nodata &    \pobs{S} &    0.09 &   \notanins &         \\ 
J135152.7$+$462149 &  0.925 &              Cluster of Galaxies &              RXC J1351.7+4622 &    \pobs{S} &    0.35 &       \unde &         \\ 
J140818.1$+$792113 &  0.890 &                          \nodata &               RX J1408.3+7921 &    \pobs{S} & $-$0.88 &       \unde &         \\ 
J141256.0$+$792204 &  0.904 &                          \nodata &               RX J1412.9+7922 &   \pobs{CS} &    0.50 &         INS &         \\ 
J141703.1$+$314249 &  0.812 &                          \nodata &               RX J1417.0+3142 &    \pobs{S} & $-$0.44 &   \notanins &         \\ 
J142644.1$+$500633 &  0.914 &                          \nodata &               RX J1426.7+5006 &    \pobs{S} & $-$0.85 &   \unde &         \\ 
J143652.6$+$582104 &  0.885 &                          \nodata &               RX J1436.8+5821 &    \pobs{S} & $-$0.31 &   \notanins &         \\ 
J144359.5$+$443124 &  0.903 &                          \nodata &               RX J1443.9+4431 &    \pobs{S} &    0.37 &   \unde     &         \\ 
J145010.6$+$655944 &  0.947 &        Cataclysmic Variable Star &               RX J1450.1+6559 &    \pobs{C} & $-$0.83 &   \notanins &         \\ 
J145234.9$+$323536 &  1.000 &                          \nodata &               RX J1452.5+3235 &    \pobs{C} &    0.00 &       \unde &         \\ 
J145729.4$+$083356 &  0.804 &                 Seyfert 1 Galaxy &               RX J1457.4+0833 &    \pobs{S} & $-$0.21 &   \notanins &         \\ 
J153840.1$+$592118 &  0.961 & LINER-type Active Galaxy Nucleus &                     NGC  5982 &    \pobs{S} &    0.73 &   \notanins &         \\ 
J160518.8$+$324907 &  1.000 &                             Star &               RX J1605.3+3249 &  \pobs{CXH} & $-$0.70 &         INS &         \\ 
J161455.4$-$252800 &  0.873 &                          \nodata &                       \nodata &   \pobs{CH} &    0.95 &   \notanins &         \\ 
J162721.6$-$244144 &  0.925 &      Variable Star of Orion Type &                  V* V2247 Oph &  \pobs{CXH} &    0.69 &   \notanins &         \\ 
J163212.8$-$244013 &  0.820 &      Variable Star of Orion Type &                  V* V2248 Oph &    \pobs{H} &    1.00 &   \notanins &         \\ 
J163421.2$+$570933 &  0.961 &          Variable of BY Dra type &                     V* CM Dra &   \pobs{XH} & $-$0.30 &   \notanins &         \\ 
J163910.7$+$565637 &  1.000 &            Active Galaxy Nucleus &               RX J1639.1+5656 &    \pobs{C} & $-$0.13 &   \notanins &         \\ 
J164020.0$+$673612 &  0.892 &                       Flare Star &                      GJ  3971 &    \pobs{S} & $-$0.12 &   \notanins &         \\ 
J165344.7$+$101159 &  0.905 &                          \nodata &               RX J1653.7+1011 &    \pobs{S} &    0.72 &       \unde &         \\ 
J171502.4$-$333344 &  0.961 &                          \nodata &                       \nodata &    \pobs{S} &    0.19 &   \unde &         \\ 
J172148.4$-$051729 &  0.839 &                          \nodata &                       \nodata &    \pobs{S} &    0.92 &       \unde &         \\ 
J173006.4$+$033813 &  0.860 &                          \nodata &                       \nodata &    \pobs{S} &    0.62 &   \notanins &         \\ 
J173157.7$-$335007 &  0.830 &            Low Mass X-ray Binary &             NAME SLOW BURSTER &  \pobs{CXS} &    1.00 &   \notanins &         \\ 
J173253.6$-$371200 &  1.000 &                          \nodata &                       \nodata &     \nodata &    0.02 &       \unde &         \\ 
J173319.3$-$255416 &  0.961 &                          \nodata &                       \nodata &   \pobs{XH} &    1.00 &   \notanins &         \\ 
J173546.9$-$302859 &  0.864 &            Low Mass X-ray Binary &                       \nodata &   \pobs{CH} &    1.00 &   \notanins &         \\ 
J180132.3$-$203132 &  0.944 &            Low Mass X-ray Binary &                     X Sgr X-3 &  \pobs{CXH} &    0.99 &   \notanins &         \\ 
J181506.1$-$120545 &  0.846 &            Low Mass X-ray Binary &                    4U 1812-12 &  \pobs{CHS} &    1.00 &   \notanins &         \\ 
J182102.0$-$161309 &  0.915 &                          \nodata &                       \nodata &   \pobs{CX} &    0.74 &       \unde &         \\ 
J183543.6$-$325928 &  0.823 &            Low Mass X-ray Binary &              NAME RX J1832-33 &  \pobs{CXH} &    0.93 &   \notanins &         \\ 
J185557.3$+$233410 &  0.820 &          Variable of BY Dra type &                   V* V775 Her &     \nodata & $-$0.11 &       \unde &         \\ 
J191426.1$+$245641 &  0.843 &        Cataclysmic Variable Star &                   V* V407 Vul & \pobs{CXHS} &    0.95 &   \notanins &         \\ 
J204249.0$+$412242 &  0.805 &                       Flare Star &                  V* V1589 Cyg &    \pobs{S} & $-$0.22 &       \unde &         \\ 
J205549.4$+$435216 &  0.857 &                          \nodata &                       \nodata &    \pobs{S} &    0.92 &   \notanins &         \\ 
J210324.7$+$193026 &  0.838 &                          \nodata &                       \nodata &    \pobs{S} &    0.62 &   \notanins &         \\ 
J212700.3$+$101108 &  0.832 &                          \nodata &                       \nodata &    \pobs{S} &    0.50 &   \unde &         \\ 
J213944.3$+$595016 &  0.821 &                          \nodata &                       \nodata &   \pobs{XS} &    0.90 &   \notanins &         \\ 
J214918.6$+$095130 &  0.884 &                          \nodata &               RX J2149.3+0951 &    \pobs{S} &    0.43 &   \notanins &         \\ 
J221144.6$-$034947 &  0.804 &                          \nodata &              2E 2209.1$-$0404 &    \pobs{C} &    0.86 &   \unde &         \\ 
J221403.0$+$124207 &  0.821 &                       Dwarf Nova &                     V* RU Peg &    \pobs{X} &    0.48 &   \notanins &         \\ 
J223832.0$-$151809 &  0.810 &                       Flare Star &                     V* EZ Aqr &     \nodata & $-$0.50 &   \notanins &         \\ 
J230334.0$+$152019 &  0.925 &                          \nodata &               RX J2303.5+1520 &    \pobs{H} &    1.00 &       \unde &         \\ 
J230340.4$-$352420 &  0.845 &                          \nodata &                       \nodata &    \pobs{S} & $-$0.95 &       \unde &         \\ 
J231543.7$-$122159 &  0.950 &          High proper-motion Star &                     LHS  3918 &    \pobs{S} & $-$0.13 &   \notanins &         \\ 
J231728.9$+$193651 &  0.904 &          Double or multiple star &  [ZEH2003] RX J2317.4+1936  3 &    \pobs{S} & $-$0.40 &   \notanins &         \\ 
J233407.2$-$033533 &  0.860 &                          \nodata &                       \nodata &     \nodata &    0.80 &       \unde &         \\ 
J233757.2$+$271031 &  0.831 &                          \nodata &               RX J2337.9+2710 &   \pobs{XH} &    1.00 &       \unde &         \\ 
J234305.1$+$363226 &  0.868 &                          \nodata &                       \nodata &    \pobs{S} & $-$0.44 &   \notanins &         \\ 
J234421.2$+$213601 &  0.876 &                          \nodata &               RX J2344.3+2136 &    \pobs{S} & $-$0.10 &   \notanins &         \\ 
J234836.5$-$273935 &  0.829 &                          \nodata &                       \nodata &    \pobs{S} & $-$0.32 &   \notanins &         \\ 
\enddata

\tablenotetext{a}{Facilities key:  \pobs{C}, \chandra; \pobs{X}, \xmm; \pobs{H}, \rosat\ HRI; \pobs{S}, \swift}

\label{tab:sources}

\end{deluxetable}


\newpage

\begin{deluxetable}{lllllllllllll}
\rotate
\tablewidth{0pt}
\tabletypesize{\footnotesize}
\tablecolumns{13}

\tablecaption{Objects Observed and Detected with \swift/XRT}

\tablehead
{
  	\colhead{1RXS J}   &
 	 \colhead{Obs. ID\tnm{a}}  &
  	\colhead{XRT J} &
  	\colhead{$R$(\arcsec)\tnm{b}}    &
  	\colhead{UVOT J}    &
  	\colhead{$P_{\rm X}$\tnm{c}} &
  	\colhead{2MASS J}    &
	\colhead{$P_{\rm X}$\tnm{d}}   &
 	 \colhead{$P_{\rm U}$\tnm{e}}  &
  	\colhead{USNO J} &
  	\colhead{$P_{\rm X}$\tnm{f}}    &
  	\colhead{$P_{\rm U}$\tnm{g}}    &
	\colhead{ID}
}

\startdata
J003854.9-034252 & 00035514001 & J003856.7-034346.7 & 61.4 & J003856.9-034345.2 & -3.1 & J003856.8-034345.0 & -2.9 & -4.5 & J003856.8-034345.1 & -2.1 & -3.0 & \notanins \\ 
J012428.1-335504 & 00035515001 & J012427.9-335508.4 & 4.8 & J012427.7-335509.5 & -3.0 & J012427.6-335508.6 & -5.2 & -6.3 & J012427.6-335508.4 & -1.6 & -2.3 & \notanins \\ 
J013653.5-351012 & 00035507002 & J013654.8-350947.4 & 29.9 & J013654.6-350947.6 & -3.5 & J013654.4-350952.4 & -1.9 & -2.2 & J013654.4-350952.1 & -1.2 & -1.5 & \notanins \\ 
J014205.0+213045 & 00035516001 & J014203.3+213123.6 & 44.9 & & & J014203.4+213117.1 & -1.7 & & & & & \notanins\tnm{h}\\ 
J015311.6-210545 & 00035510001 & J015311.4-210542.8 & 3.3 & J015311.4-210543.0 & -3.8 & J015311.3-210543.2 & -6.0 & -5.7 & J015311.3-210543.2 & -2.8 & -2.4 & \notanins \\ 
J020146.5+011717 & 00035495001 & J020146.8+011703.6 & 14.3 & J020146.9+011705.7 & -3.0 & J020146.9+011705.8 & -5.4 & -6.6 & J020146.7+011658.5 & -1.3 & -1.3 & \notanins \\ 
J022152.6+281047 & 00035519002 & J022153.2+281108.9 & 23.5 & J022153.4+281114.1 & -2.3 & J022153.3+281114.4 & -4.0 & -5.3 & J022153.3+281114.5 & -1.0 & -2.5 & \notanins \\ 
J024814.8-193954 & 00035520002 & J024817.8-194000.2 & 43.6 & J024817.8-193957.3 & -3.3 & J024817.8-193956.9 & -2.7 & -5.0 & J024817.8-193957.0 & -1.8 & -4.7 & \notanins \\ 
J024946.0-382540 & 00035502002 & J024946.1-382535.1 & 5.0 & J024945.9-382537.5 & -3.6 & J024945.9-382536.6 & -5.6 & -6.2 & J024945.8-382537.5 & -1.7 & -3.2 & \notanins \\ 
J031413.7-223533 & 00031180001 & J031413.4-223542.2 & 9.8 & J031413.5-223538.7 & -2.8 & & & & & & & \notanins \\ 
J032620.8+113106 & 00035503002 & J032621.0+113050.2 & 16.2 & J032620.9+113050.5 & -3.8 & J032621.2+113054.1 & -4.3 & -4.2 & J032621.2+113056.3 & -1.3 & -1.3 & \notanins \\ 
J035101.5+141404 & 00035522001 & J035101.1+141341.9 & 22.7 & J035100.8+141339.5 & -3.1 & J035100.7+141339.7 & -4.4 & -6.2 & J035100.7+141340.0 & -1.2 & -3.1 & \notanins \\ 
J040314.6-360927 & 00035523001 & J040316.1-360933.7 & 20.1 & J040315.9-360934.9 & -2.8 & J040315.8-360936.2 & -4.6 & -5.3 & J040315.8-360936.4 & -1.4 & -2.0 & \notanins \\ 
J040814.3+400723 & 00035525002 & J040814.1+400734.7 & 11.8 & J040814.0+400731.8 & -3.3 & J040814.0+400732.0 & -4.4 & -5.8 & J040814.1+400732.2 & -1.7 & -2.8 & \notanins \\ 
J040817.9+294951 & 00035526003 & J040817.8+294938.8 & 12.2 & & & J040818.3+294931.1 & -3.0 & & & & & \notanins \\ 
J040913.8+110833 & 00035527001 & J040911.8+110835.7 & 28.8 & & & J040911.8+110827.8 & -1.5 & & & & & \notanins\tnm{h} \\ 
J041215.8+644407 & 00035506003 & J041216.8+644350.5 & 17.9 & J041217.0+644347.9 & -3.3 & J041216.9+644355.9 & -4.2 & -3.8 & J041216.3+644353.0 & -1.1 & -0.8 & \notanins \\ 
J042513.4+171548 & 00035528001 & J042513.5+171606.2 & 18.4 & J042513.5+171605.4 & -3.2 & J042513.5+171605.5 & -5.8 & -5.9 & J042513.5+171605.5 & -2.7 & -3.0 & \notanins \\ 
J043334.8+204437 & 00035504001 & J043334.4+204446.2 & 10.4 & J043334.1+204443.6 & -3.4 & J043333.9+204446.1 & -4.2 & -4.8 & J043333.9+204446.0 & -1.0 & -1.5 & \notanins \\ 
J044048.0+292440 & 00035529001 & J044048.5+292434.5 & 8.6 & & & J044048.3+292433.9 & -2.2 & & & & & \notanins \\ 
J050909.9+152740 & 00035530001 & J050910.2+152728.6 & 12.2 & J050910.0+152729.0 & -2.9 & J050909.9+152732.3 & -4.4 & -4.8 & J050910.1+152728.9 & -2.2 & -2.1 & \notanins \\ 
J051028.9+022051 & 00035531001 & J051029.5+022058.1 & 11.5 & & & & & & J051029.4+022058.3 & -2.4 & & \notanins \\ 
J051315.9+025227 & 00035532001 & J051318.9+025147.0 & 60.5 & & & & & & J051317.8+025143.6 & -0.2 & & \notanins\tnm{i} \\ 
J051319.1+013525 & 00035511001 & J051319.1+013448.3 & 36.6 & J051318.9+013446.8 & -3.6 & J051319.0+013446.9 & -5.4 & -6.3 & J051319.0+013447.0 & -1.7 & -4.4 & \notanins \\ 
J051354.0+023722 & 00035533002 & J051402.0+023906.6 & 160.1 & J051401.9+023910.7 & -2.9 & J051401.9+023910.9 & -4.1 & -6.2 & J051402.0+023911.0 & -1.0 & -3.1 & \notanins \\ 
J051723.3-352152 & 00035494002 & J051722.7-352154.2 & 6.9 & J051722.7-352155.4 & -3.3 & J051722.9-352154.5 & -6.2 & -6.2 & J051722.9-352154.1 & -2.0 & -1.9 & \notanins \\ 
J051854.5+323827 & 00035534004 & J051855.2+323828.4 & 9.6 & & & J051855.0+323831.2 & -4.1 & & & & & \notanins \\ 
J052929.6+031838 & 00035535001 & J052929.6+031821.2 & 16.7 & J052929.4+031820.8 & -2.9 & J052929.4+031820.4 & -4.6 & -6.0 & J052929.4+031820.5 & -1.4 & -3.4 & \notanins \\ 
J062203.6+744106 & 00035536001 & J062207.7+744041.2 & 29.6 & J062207.0+744035.9 & -2.7 & J062207.0+744035.9 & -4.2 & -7.2 & J062207.1+744036.1 & -0.8 & -3.4 & \notanins \\ 
J075523.9+372634 & 00035537001 & J075523.2+372620.1 & 16.1 & J075523.1+372618.9 & -2.4 & & & & J075523.1+372618.8 & -2.1 & -3.4 & \notanins \\ 
J075556.7+832310 & 00035499002 & J075552.2+832302.1 & 11.0 & J075552.9+832301.1 & -2.9 & J075553.9+832304.8 & -4.9 & -4.9 & J075553.9+832305.1 & -1.5 & -1.5 & \notanins \\ 
J082124.5-362848 & 00035538001 & J082126.9-362647.9 & 123.7 & & & J082126.9-362646.8 & -6.1 & & & & & \notanins \\ 
J082355.1+394745 & 00035539001 & J082355.4+394747.0 & 4.8 & & & & & & J082355.6+394747.6 & -2.1 & & \notanins \\ 
J084127.7-102843 & 00035540001 & J084127.4-102835.1 & 8.7 & & & J084127.2-102835.6 & -1.9 & & & & & \unde \\ 
J090137.2-052341 & 00035541001 & J090137.0-052414.2 & 33.3 & J090137.4-052409.1 & -2.4 & J090137.4-052409.5 & -1.4 & -4.5 & J090137.4-052409.2 & -0.8 & -5.0 & \notanins \\ 
J090717.4+225254 & 00035509002 & J090718.0+225221.6 & 33.7 & J090718.0+225221.9 & -3.8 & J090718.0+225221.2 & -6.2 & -6.5 & J090718.0+225221.5 & -4.8 & -3.2 & \notanins \\ 
J091010.2+481317 & 00035496001 & J091010.0+481338.4 & 21.5 & & & J091010.0+481341.5 & -2.8 & & & & & \notanins \\ 
J094432.8+573544 & 00036709001 & J094432.1+573531.5 & 13.4 & J094432.3+573536.0 & -2.2 & & & & J094432.3+573536.5 & -1.1 & -2.6 & \notanins \\ 
J094454.2-122047 & 00035512002 & J094453.9-122056.1 & 9.7 & J094454.0-122053.2 & -3.1 & J094454.2-122054.3 & -5.5 & -5.9 & J094454.2-122054.7 & -1.4 & -1.8 & \notanins \\ 
J102213.7+735433 & 00035543001 & J102213.6+735436.6 & 3.7 & J102213.4+735433.9 & -2.9 & & & & J102213.5+735433.6 & -1.8 & -3.8 & \notanins \\ 
J110521.6-073525 & 00035544001 & J110522.3-073558.3 & 35.0 & J110522.0-073559.1 & -3.5 & J110522.0-073558.7 & -5.0 & -6.2 & J110522.0-073558.8 & -1.4 & -3.5 & \notanins \\ 
J112430.5+435131 & 00035545003 & J112429.7+435123.6 & 10.8 & J112429.7+435125.2 & -2.0 & & & & J112429.6+435125.5 & -2.0 & -3.3 & \notanins \\ 
J120711.0+364745 & 00035546001 & J120716.7+364759.7 & 70.3 & J120717.7+364800.5 & -2.7 & J120717.8+364800.7 & -1.4 & -4.2 & J120717.6+364758.4 & -0.6 & -2.5 & \notanins \\ 
J121732.6+152843 & 00035547001 & J121733.0+152850.7 & 9.7 & J121732.7+152844.8 & -3.2 & J121732.7+152844.6 & -2.0 & -5.8 & J121732.7+152844.4 & -1.2 & -3.8 & \notanins \\ 
J121900.7+110727 & 00035548001 & J121858.9+110733.6 & 27.0 & J121858.8+110734.8 & -3.8 & J121859.3+110733.8 & -4.8 & -4.7 & J121859.1+110736.8 & -1.4 & -1.5 & \notanins \\ 
J122308.4+110054 & 00035549001 & J122307.9+110037.4 & 17.6 & J122307.2+110038.3 & -2.0 & & & & J122307.2+110038.4 & -0.8 & -4.6 & \notanins \\ 
J124849.0+333454 & 00035550001 & J124851.5+333506.0 & 34.1 & J124851.8+333505.6 & -2.9 & & & & J124851.8+333506.0 & -1.7 & -3.5 & \notanins \\ 
J125015.2+192357 & 00035500003 & J125014.7+192348.5 & 10.3 & J125014.9+192350.5 & -2.8 & & & & J125014.9+192350.6 & -1.5 & -3.3 & \notanins \\ 
J125947.9+275636 & 00035176002 & J125935.9+275730.5 & 166.8 & J125935.5+275733.7 & -2.9 & J125935.7+275734.5 & -3.5 & -5.1 & J125935.7+275734.0 & -1.0 & -1.9 & \notanins \\ 
J130034.2+054111 & 00035501001 & J130032.8+054107.6 & 20.1 & & & J130033.4+054108.0 & -4.2 & & & & & \notanins \\ 
J130205.2+155122 & 00035552001 & J130205.2+155134.0 & 12.1 & & & & & & J130204.8+155137.8 & -1.1 & & \unde \\ 
J130402.8+353316 & 00035505003 & J130402.8+353312.5 & 3.5 & J130402.8+353316.8 & -2.9 & J130402.7+353316.8 & -2.4 & -4.4 & J130402.7+353316.4 & -1.4 & -2.5 & \notanins \\ 
J130631.3+192229 & 00035553002 & J130631.1+192241.9 & 13.2 & J130630.9+192244.4 & -3.0 & J130630.9+192243.8 & -2.6 & -4.6 & J130631.0+192243.2 & -2.4 & -3.6 & \notanins \\ 
J131011.9+474521 & 00035554003 & J131012.0+474515.7 & 5.5 & J131012.1+474513.7 & -3.2 & J131012.6+474519.0 & -4.6 & -4.5 & J131012.6+474519.0 & -1.1 & -1.0 & \notanins \\ 
J133032.3+720931 & 00035556002 & J133031.8+720928.0 & 3.5 & J133031.2+720925.7 & -3.0 & & & & J133031.3+720925.7 & -1.7 & -3.2 & \notanins \\ 
J133825.0-251634 & 00035557001 & J133825.3-251648.2 & 15.0 & J133825.4-251647.0 & -3.4 & J133825.6-251646.5 & -5.5 & -6.1 & J133825.7-251646.2 & -1.1 & -1.5 & \notanins \\ 
J134619.1-141834 & 00035558001 & J134619.1-141855.2 & 21.2 & J134619.2-141856.9 & -2.8 & J134619.2-141856.4 & -2.6 & -4.5 & J134619.2-141856.5 & -1.8 & -3.4 & \notanins \\ 
J141256.0+792204 & 00035558001 & J134619.1-141855.2 & 21.2 & & & & & & & & & INS\tnm{j} \\ 
 J141703.1+314249 & 00035562001 & J141702.6+314243.9 & 7.9 & J141702.6+314246.0 & -3.1 & J141702.9+314247.1 & -5.0 & -5.2 & J141702.8+314247.1 & -1.3 & -1.6 & \notanins \\ 
J142644.1+500633 & 00035563001 & J142647.5+500906.6 & 157.1 & J142648.1+500908.9 & -3.2 & & & & & & & \notanins \\ 
J143652.6+582104 & 00035564002 & J143653.9+582051.7 & 15.9 & J143653.5+582049.9 & -2.9 & J143653.0+582054.8 & -5.0 & -5.2 & J143654.5+582050.6 & -1.1 & -1.0 & \notanins \\ 
J144359.5+443124 & 00035565001 & J144400.5+443124.2 & 11.5 & J144400.3+443118.7 & -1.9 & & & & J144400.5+443119.5 & -1.1 & -3.0 & \unde \\ 
J145729.4+083356 & 00035566001 & J145728.9+083427.6 & 32.5 & J145728.9+083423.0 & -2.9 & J145728.9+083422.5 & -2.1 & -4.5 & J145729.0+083422.5 & -1.0 & -2.3 & \notanins \\ 
J153840.1+592118 & 00035567001 & J153839.9+592120.2 & 2.5 & J153839.9+592121.0 & -3.8 & J153839.7+592120.9 & -5.3 & -5.4 & J153840.0+592120.9 & -2.9 & -4.6 & \notanins \\ 
J164020.0+673612 & 00035568001 & J164020.9+673603.8 & 9.7 & J164020.3+673607.3 & -2.1 & J164020.6+673604.6 & -5.8 & -5.1 & J164020.6+673605.0 & -1.9 & -1.4 & \notanins \\ 
J171502.4-333344 & 00035508001 & J171501.6-333335.8 & 12.2 & & & J171501.8-333333.2 & -1.3 & & & & & \unde \\ 
J173006.4+033813 & 00035571001 & J173006.1+033818.8 & 6.6 & J173006.4+033819.5 & -3.0 & & & & J173006.4+033819.9 & -0.9 & -2.7 & \notanins \\ 
J205549.4+435216 & 00035574001 & J205551.2+435224.5 & 21.8 & & & J205551.2+435224.5 & -6.5 & & & & & \notanins \\ 
J210324.7+193026 & 00035575002 & J210323.2+193054.4 & 34.7 & J210323.2+193055.7 & -2.8 & J210323.2+193055.6 & -5.1 & -6.2 & J210323.2+193055.7 & -2.0 & -4.3 & \notanins \\ 
J212700.3+101108 & 00035576001 & J212700.3+101122.2 & 14.3 & J212700.1+101119.0 & -1.7 & & & & J212700.1+101119.1 & -1.3 & -4.5 & \unde \\ 
J213944.3+595016 & 00035577001 & J213944.3+595017.1 & 1.1 & & & J213944.9+595014.9 & -1.3 & & & & & \unde \\ 
J214918.6+095130 & 00035578002 & J214919.5+095138.6 & 16.2 & & & J214919.6+095137.1 & -2.4 & & & & & \notanins \\ 
J231543.7-122159 & 00035498001 & J231544.1-122150.6 & 10.4 & J231544.0-122150.2 & -3.1 & J231543.7-122148.4 & -5.2 & -5.5 & J231543.7-122149.0 & -1.2 & -1.6 & \notanins \\ 
J231728.9+193651 & 00035497001 & J231728.2+193646.5 & 9.7 & J231728.3+193646.0 & -3.3 & J231728.0+193646.9 & -5.8 & -5.7 & J231728.1+193646.9 & -1.7 & -1.6 & \notanins \\ 
J234305.1+363226 & 00035581001 & J234307.1+363217.7 & 26.2 & & & J234306.2+363213.1 & -4.0 & & & & & \notanins \\ 
J234421.2+213601 & 00035582001 & J234421.0+213602.8 & 2.5 & J234421.1+213605.6 & -3.0 & J234420.8+213605.0 & -5.1 & -5.1 & J234420.8+213605.2 & -1.6 & -1.6 & \notanins \\ 
J234836.5-273935 & 00035583003 & J234835.7-273938.2 & 10.5 & J234835.7-273940.6 & -3.1 & J234836.1-273938.4 & -4.7 & -4.7 & J234836.0-273938.5 & -1.2 & -1.1 & \notanins 
\enddata

\tablenotetext{a}{\swift/XRT \obsid}
\tablenotetext{b}{Separation $R$ between RASS/BSC source and \swift\ source}
\tablenotetext{c}{Log probability of chance association between UVOT and XRT source}
\tablenotetext{d}{Log probability of chance association between 2MASS and XRT source}
\tablenotetext{e}{Log probability of chance association between 2MASS and UVOT source}
\tablenotetext{f}{Log probability of chance association between USNO and XRT source}
\tablenotetext{g}{Log probability of chance association between USNO and UVOT source}
\tablenotetext{h}{XRT source found to be extended}
\tablenotetext{i}{XRT source saturated}
\tablenotetext{j}{During the process of this work, identified as an INS (Calvera) \citep{rutledge08}.}

\label{tab:swift}

\end{deluxetable}

 \newpage

\begin{deluxetable}{llll}
\tablewidth{0pt}
\tabletypesize{\small}
\tablecolumns{4}
\label{tab:sources}

\tablecaption{Objects Undetected with \swift/XRT}

\tablehead{
  \colhead{1RXSJ}   &
  \colhead{Observation ID}  &
  \colhead{Dur.\tnm{a}}    &
  \colhead{Flux Limit\tnm{b}}   
 }

\startdata
J020210.5-020616	&	00035517001	&	0.9	&	1.8\tee{-13}	\\
J020503.6-173717	&	00035518002	&	1.2	&	1.3\tee{-13}	\\
J033642.5-095506	&	00035521001	&	1.2	&	1.5\tee{-13}	\\
J040543.6-282114	&	00035524001	&	2.5	&	2.5\tee{-13}	\\
J094831.1-333814	&	00035542001	&	1.2	&	1.9\tee{-13}	\\
J122940.6+181645	&	00036885001	&	0.9	&	1.7\tee{-13}	\\
J125721.8+365431	&	00035551001	&	1.0	&	1.4\tee{-13} 	\\
J132041.2-030010	&	Multiple	&		&	5.5\tee{-13} 	\\
J135152.7+462149	&	00035559001	&	1.2	&	1.3\tee{-13}	\\
J140818.1+792113	&	00035560002	&	1.0	&	1.3\tee{-13}	\\
J142644.1+500633	&	00035563001	&	0.9	&	7.4\tee{-13} 	\\
J165344.7+101159	&	00035569001	&	1.4	&	1.3\tee{-13}	\\
J172148.4-051729	&	00035570001	&	1.5	&	1.5\tee{-13}	\\
J204249.0+412242	&	00035573001	&	1.0	&	5.1\tee{-13}	\\
J230340.4-352420	&	Multiple	&		&	6.2\tee{-13} 
\enddata

\tablenotetext{a}{Observation duration (ks)}
\tablenotetext{a}{in \cgsflux (0.1-2.4 keV)}

\label{tab:nonswift}

\end{deluxetable} 

\newpage 

\begin{deluxetable}{llll}
\tablewidth{0pt}
\tabletypesize{\small}
\tablecolumns{3}
\label{tab:sources}

\tablecaption{Full-Sky Upper Limits on the Number of INSs in the RASS/BSC}

\tablehead{
  \colhead{INS Candidate Restrictions}   &
  \colhead{90\% Confidence}  &
  \colhead{99\% Confidence}    
 }

\startdata
None				&	46 	&	66	\\
Non-variable			&	37	&	54	\\
Non variable, spectrally soft	&	31	&	46	\\
R03				&	67	&	104
\enddata

\label{tab:upperlimit}

\end{deluxetable} 


\newpage

\begin{deluxetable}{llc}
\tablewidth{0cm}
\tablecaption{Isolated Neutron Star Candidates Detected in
  Second-Epoch Observations\label{tab:ins} }
\tablecolumns{3}
\tablehead{
\colhead{1RXSJ} &
\colhead{\pnoid} &
\colhead{Flux (\tee{-13} \cgsflux)}
}
\startdata
J044048.0+292440	&	0.914	&	36.6		\\ analysis. 
J084127.7$-$102843 	&	0.836	&	11		\\ 
J130205.2+155122 	&	1.000	&	3		\\  
J144359.5+443124	&	0.903	&	8.5		\\
J171502.4$-$333344 	&	0.961	&	32.5		\\ 
J212700.3+101108 	&	0.832	&	15.1		\\  
J213944.3+595016 	&	0.821	&	45		\\ 
J230334.0+152019 	&	0.925	&	6.9		\\ 
J233757.2+271031 	&	0.831	&	4.5		\\  
\hline
\enddata

\end{deluxetable}


\end{document}